\documentclass{aa}
\usepackage{epsf}

\def\ave#1{\langle #1 \rangle}

\def\rund#1{\left( #1 \right)}
\def \la {\mathrel{\vcenter
     {\offinterlineskip \hbox{$<$}\hbox{$\sim$}}}}
\def \ga {\mathrel{\vcenter
     {\offinterlineskip \hbox{$>$}\hbox{$\sim$}}}}
\begin{document}

%\thesaurus{06(08.14.1, 08.02.1, 02.08.1, 02.07.2, 02.14.1, 02.05.1)}

\title{Coalescing neutron stars -- a step towards physical models}
\subtitle{III. Improved numerics and different neutron star masses and spins}

\author{M.~Ruffert\inst{1}\thanks{e-mail: {\tt m.ruffert@ed.ac.uk}}
\and H.-Th.~Janka\inst{2}\thanks{e-mail: {\tt thj@mpa-garching.mpg.de}}
}
\institute{Department of Mathematics \& Statistics, University of Edinburgh,
Edinburgh, EH9 3JZ, Scotland, U.K.
\and Max-Planck-Institut f\"ur Astrophysik, Postfach 1317, 85741
Garching, Germany}
\offprints{H.-Th.~Janka}

%\date{Received; accepted}

\abstract{ 
In this paper we present a compilation of results from our most advanced 
neutron star merger simulations. Special aspects of these models were 
refered to in earlier publications (Ruffert \& Janka 1999, Janka et al.~1999),
but a description of the employed numerical procedures and a more complete 
overview over a large number of computed models are given here.
The three-dimensional hydrodynamic simulations were done with a 
code based on the Piecewise Parabolic Method (PPM), which solves 
the discretized conservation laws for mass, momentum, energy and, in
addition, for the electron lepton number in an Eulerian frame of reference.
Up to five levels of nested cartesian grids ensure higher numerical resolution 
(about 0.6$\,$km) around the center of mass while the evolution is followed
in a large computational volume (side length between 300 and 400$\,$km). 
The simulations are basically Newtonian, but gravitational-wave emission and
the corresponding back-reaction on the hydrodynamic flow are taken into account.
The use of a physical nuclear equation of state allows us to follow the thermodynamic
history of the stellar medium and to compute the energy and lepton number loss 
due to the emission of neutrinos.
The computed models differ concerning the neutron star masses and mass ratios,
the neutron star spins, the numerical resolution expressed by the cell size of 
the finest grid and the number of grid levels, and the calculation of the 
temperature from the solution of the entropy equation instead of the energy
equation. The models were evaluated for the corresponding gravitational-wave
and neutrino emission and the mass loss which occurs during the dynamical phase
of the merging. The results can serve for comparison with smoothed particle
hydrodynamics (SPH) simulations. In addition, they define a reference point
for future models with a better treatment of general relativity and with
improvements of the complex input physics.
Our simulations show that the details of the gravitational-wave emission are
still sensitive to the numerical resolution, even in our highest-quality
calculations. The amount of mass which can be ejected from neutron star 
mergers depends strongly on the angular momentum of the system. Our results
do not support the initial conditions of temperature and proton-to-nucleon 
ratio needed according to recent work for producing a solar r-process
pattern for nuclei around and above the $A\approx 130$ peak. The improved models
confirm our previous conclusion that gamma-ray bursts are not powered by 
neutrino emission during the dynamical phase of the merging of two neutron stars.
}

\maketitle

\section{Introduction}
\label{introduction}
The binary pulsar PSR~1913+16 (Hulse \& Taylor 1975) is the most 
famous example for a binary
system containing two neutron stars, among another $\sim\,1000$ of such systems
expected to exist in our Galaxy. High-precision measurements show that the change
in time of the orbital parameters of PSR~1913+16 is consistent with expectations
from the theory of general relativity, which predicts the emission of gravitational
waves and a continuous decrease of the orbital separation. Therefore, these systems
have a finite lifetime of typically hundreds of millions up to billions of years.
As the two stars spiral in towards each other, the evolution accelerates because
the gravitational-wave emission rises strongly with decreasing distance. When
the orbital separation has shrunk to only a few stellar radii, the system
has become a strong source of gravitational waves with a frequency around 
100$\,$Hz. It will end its life within milliseconds in the final, catastrophic
merging of the two neutron stars, emitting a powerful outburst of gravitational 
radiation which carries important information about the properties of the 
merging stars, the dynamics of the coalescence, and the remnant left behind.

With an estimated rate of about 10$^{-5}$ events per year per galaxy (e.g.,
see the recent numbers in Bulik et al.~1999,
Fryer et al.~1999, Kalogera \& Lorimer 2000,
and references therein) neutron star mergers are among the most frequent and
most promising candidates for gravitational-wave emission which is strong enough 
to be measurable by the upcoming interferometric experiments in the U.S. (LIGO),
Europe (GEO600, VIRGO), and Japan (TAMA) (Thorne 1995). Theoretical models
and wave templates, however, are needed to help filter out the weak signals
from disturbing background noise. Gravitational waves from neutron
star mergers could be one of the most fruitful ways to learn about the internal
properties of neutron stars. 

Merging neutron stars are also considered as
possible sources of at least the subclass of short and hard cosmic gamma-ray bursts, 
especially if the merger remnant collapses to a black hole on a dynamical timescale
(for recent discussions and model calculations, see, e.g., 
Popham et al.~1999, Ruffert \& Janka 1999). Coincident detections
of gravitational waves and gamma rays would be a convincing observational confirmation
of this hypothesis and might in fact be the only possibility to identify the
central engine of a gamma-ray burst unequivocally. The X-ray satellite HETE-2,
which was launched in Fall~2000, is hoped to bring a similar breakthrough
in the observation of short bursts as the BeppoSAX satellite did in case
of the long ones.

The energy of the relativistically
expanding fireball or jet, which finally produces the observable gamma-ray burst, can
be provided by the annihilation of neutrino-antineutrino pairs (Paczy\'nski 1991, 
M\'esz\'aros \& Rees 1992, Woosley 1993a) or possibly by
magnetohydrodynamical processes (Blandford \& Znajek 1977, M\'esz\'aros and Rees 1997). 
In the former case, the gravitational
binding energy of accreted disk matter is tapped, in the latter
case the rotational energy of the central black hole could be converted into kinetic
energy of the outflow. If neutrino processes are supposed to power the gamma-ray 
burst phenomenon, very high neutrino luminosities are needed, of magnitude similar
as those from core-collapse supernovae. The rate of neutron star mergers, 
however, is much smaller (by a factor of 100--10000) than the Galactic supernova
rate. This practically excludes them as detectable sources of thermal neutrinos
in the MeV energy range, because the signals are too faint to be
measurable from extragalactic distances. Dissipative processes in 
the relativistic outflow, which are considered to produce the gamma-ray burst,
may also lead to the generation of high-energy or even ultra high-energy neutrinos 
(Paczy\'nski \& Xu 1994, Waxman \& Bahcall 1997, 2000). 
Such neutrinos might be seen in future km$^2$-scale
experiments like ICECUBE, which is currently under construction in the Antarctica.  
However, they do not carry much specific information about the origin of the 
relativistically moving particles and it is therefore not very likely that they
can yield much evidence about the nature of the central engine that powers the
gamma-ray burst.  

Neutron-rich matter, which
is ejected from the system during the dynamical phase of the merging, was 
suggested as a possible site for the rapid neutron capture process (r-process)
to produce heavy nuclei beyond the iron group
(Lattimer et al.~1974, 1976; Hilf et al.~1974;  
Eichler et al.~1989; Meyer 1989). This problem has gained new interest recently
(Rosswog et al.~1999, 2000; Freiburghaus et al.~2000).
The possible contribution to the Galactic r-process material is estimated
from the gas mass that gets unbound during the violent last stages
of the coalescence. The nuclear reactions in decompressed neutron star
matter depend sensitively on the initial conditions (neutron excess, composition, 
temperature,
density), the dynamical and, in particular, thermal history of the material, and 
the influence of beta-decays and corresponding neutrino losses. All of these 
issues are so far not well under control in theoretical models, and therefore 
hydrodynamic simulations of neutron star mergers have not (yet?) been able to 
yield conclusive results.

These questions have been the motivation for a large number of investigations of
the spiral-in phase and the ultimate merging of neutron stars. Analytic studies
and ellipsoidal treatments
concentrated on the effects of viscous dissipation for the heating and the
rotation of the stars 
(Kochanek 1992, Bildsten \& Cutler 1992, Lai 1994), the final
instability of the mass transfer near the tidal radius (e.g., 
Bildsten \& Cutler 1992; Lai et al.~1994a,b; Taniguchi \& Nakamura 1996;
Lai \& Wiseman 1996; Lombardi et al.~1997; Baumgarte 2001) and the
deformed equilibrium structure and tidal lag of the binary configuration 
prior to the dynamical interaction (Lai \& Shapiro 1995).
Hydrodynamical simulations of the coalescence were performed for Newtonian
gravity with SPH codes (e.g., Rasio \& Shapiro 1992, 1994, 1995; 
Centrella \& McMillan 1993; Zhuge et al.~1994, 1996; 
Davies et al.~1994; Rosswog et al.~1999, 2000) and with grid-based methods
(e.g., Oohara \& Nakamura 1990, Nakamura \& Oohara 1991), 
partly including special treatments
of the gravitational-wave emission and their back-reaction on the flow by 
adding the corresponding post-Newtonian terms to the equations of hydrodynamics
(e.g., Ruffert et al.~1996, 1997a; Ruffert et al.~1997b).
More recently progress has been achieved in a wider use of the
post-Newtonian approximation (Shibata et al.~1998, 
Ayal et al. 2001, Faber \& Rasio 2000, Faber et al.~2001) and considerable
advances were made towards general 
relativistic treatments (Oohara \& Nakamura 1999, 
Shibata 1999; Shibata \& Ury\=u 2000, 2001). 

A spectacular result was obtained
by Mathews \& Wilson (1997, and references therein)
who found that relativistic effects lead to a compression
of the two neutron stars during the late stages of the spiral-in and therefore 
to their gravitational collapse to black holes
prior to the merging. This effect contradicts Newtonian models where 
tidal stretching reduces the density of the stars as they get closer.
Analytic considerations confirm the Newtonian behavior also for
the post-Newtonian case (Thorne~1998; Baumgarte et al.~1998a,b),
and more recent simulations by the Wilson group (Marronetti et al.~1999)
as well as general relativistic hydrodynamic models
by other groups (Bonazzola et al.~1999, Shibata et al.~1998) were not able to
reproduce the result of Mathews \& Wilson (1997). The latter was recognized to 
be due to an error in the approximation scheme to full general relativity
(Flanagan~1999). In any case, pre-merging
collapse of the neutron stars is a speculative option only if the nuclear
equation of state is extraordinarily soft and the neutron stars are already
very close to the maximum mass for stable single neutron stars.

The majority of the simulations by other groups
was done with simple microphysics, in particular
with a polytropic law $P = K\rho^\Gamma$ for the equation of state (EoS)
of the neutron star matter.
This is a fair approach when one is mainly interested in the calculation of the
gravitational-wave emission, which is associated with the motion of the
bulk of the mass. It offers the advantage that the influence of the stiffness
of the EoS, which determines the mass-radius relation of the neutron stars
and the amount of compression which occurs during the final plunge, can be
easily studied by choosing different values for the adiabatic index $\Gamma$.

Several years ago we started to compute merger models with a more elaborate
treatment of the EoS of the neutron star matter, using the physical description 
by Lattimer \& Swesty (1991), which enabled us to follow the thermodynamics
of the gas and to include a treatment of the neutrino production and emission
from the heated neutron stars (Ruffert et al.~1996, 1997a; Ruffert \& Janka 1998,
1999; Janka et al.~1999). Our main aims were the investigation of the relevance
for gamma-ray burst scenarios, in particular for those where the neutrino emission
had been suggested to provide the energy for the relativistic gamma-ray burst
fireball via neutrino-antineutrino
annihilation. Also the amount of mass ejection during the dynamical interaction 
and the properties of the ejected matter depend on the EoS, which cannot be 
descibed by one simple polytropic law in both the low-density and high-density
regimes. 

After publication of our first papers (Ruffert et al.~1996, 1997a), we changed
our code considerably and, in particular, we improved many features which 
had influence on the results of our simulations. For example,
we introduced nested grids to get a higher resolution of the neutron stars and 
at the same time to use a larger computational volume. In addition, we extended
the EoS table to higher temperatures and lower densities. The latter allowed us to 
reduce the density of the dilute medium that has to be assumed around the neutron 
stars on the Eulerian grid. Since the heat capacity of cold, degenerate matter 
is very small, minor numerical noise in the internal energy had induced larger 
errors in the temperature. We therefore also implemented an entropy equation,
because the entropy is numerically less problematic for calculating the temperature.
Besides these improvements, we also covered a wider range of scenarios, e.g., 
added models with opposite directions of the neutron star spins and with 
different neutron star masses as well as different mass ratios. 

All of our later publications referred to data of models which were computed with
the improved version of the code. So did the simulations of the black hole 
accretion in Ruffert \& Janka (1999) start from an initial model of the new
generation of calculations, and also in the tables of Janka et al.~(1999)
data of new neutron star merger models were listed.
So far, however, we published only very specific aspects of these new models and
did not present our results in detail. This is the purpose of the present 
publication.

In Sect.~\ref{sec:numerics} we will give a technical description of the 
code changes and improvements, 
in Sect.~\ref{sec:simulations} a list of
computed models,
in Sect.~\ref{sec:results} we shall present
the main results for the new models, and in Sect.~\ref{sec:conclusions} we shall
discuss the implications and draw conclusions.

%************************************** fig.1 ************************************
\begin{figure*}[t!]
 \begin{tabular}{cc}
  \epsfxsize=14.0cm \epsfclipon \epsffile{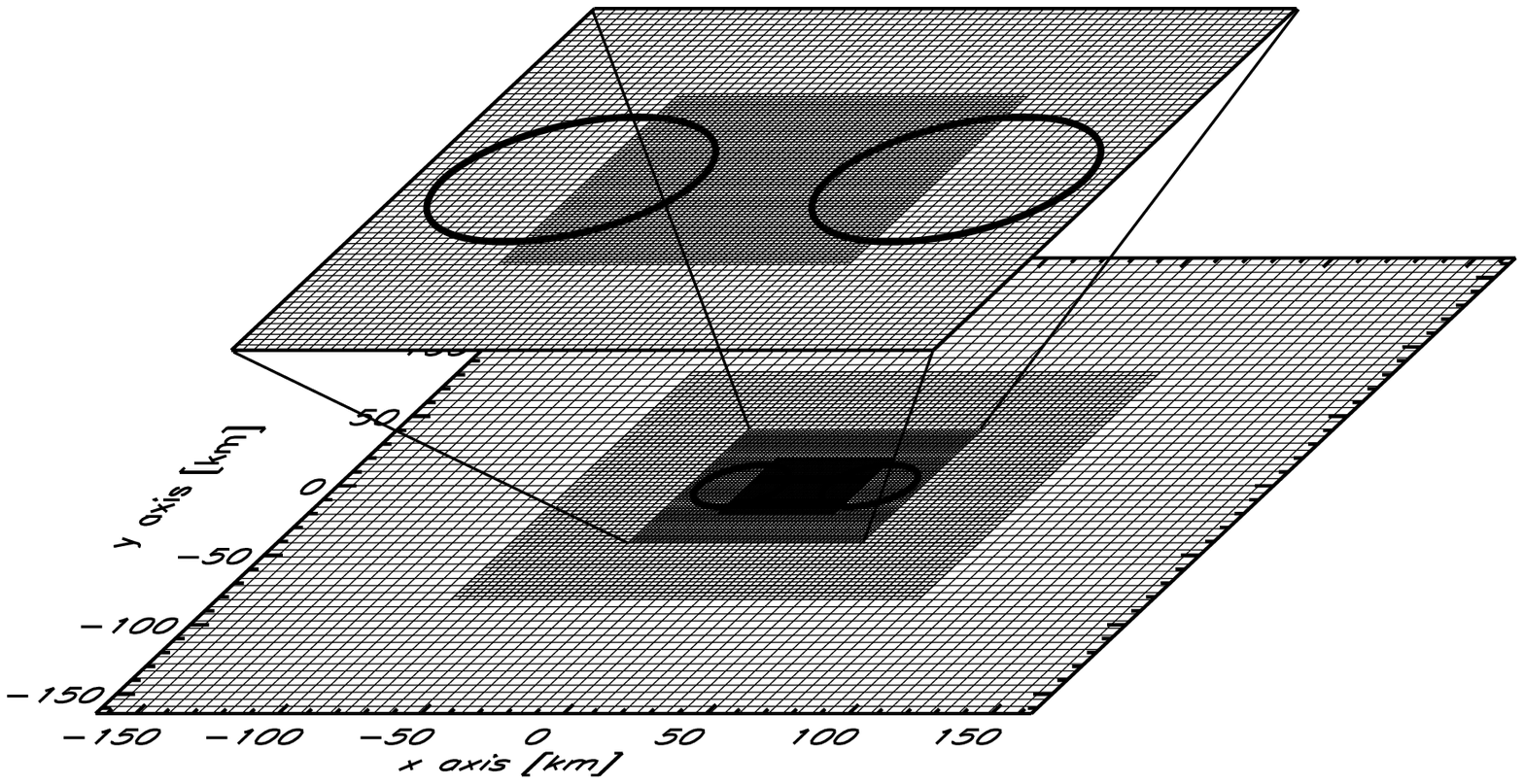} &
\raisebox{8cm}{\parbox[t]{3.3cm}{
\caption[]{\label{fig:1}
Illustration of four levels of nested grids in two (instead of three) 
spatial dimensions, each level with 64 cells in every direction,
covering a computational volume with side length of 328$\,$km. 
The innermost two grid levels are enlarged and the initial positions
of the two neutron stars in the orbital plane are indicated
}}}
%\picplace{20cm}
 \end{tabular}
\end{figure*}
%************************************** fig.1 ************************************

\section{Changes and improvements of the numerics}
\label{sec:numerics}

The three-dimensional Eulerian hydrodynamical conservation laws for mass, momentum
and energy for non-viscous flow are integrated explicitly in time on a cartesian
grid, using a finite-volume scheme which is based on the
Piecewise Parabolic Method (PPM) of Colella \& Woodward (1984). The code is
basically Newtonian, but includes the post-Newtonian terms that account for the 
local effects of gravitational-wave emission 
(the volume integral of these
local terms reproduces the quadrupole approximation)
and the corresponding back-reaction on the hydrodynamical flow according to
the formulation by Blanchet et al.~(1990) (for details, see Ruffert et al.~1996).
The Poisson equations for the gravitational potential and two additional 
potentials used in the back-reaction terms are solved by fast Fourier transforms.
For the described neutron star merger simulations, the thermodynamics of the 
stellar medium are described by a tabular version of the Lattimer \& Swesty 
(1991) EoS, which allows us to take into account the source terms for energy
and lepton number loss due to neutrino production. These source terms are
calculated with a neutrino trapping scheme which evaluates the production rates
of neutrinos and antineutrinos of all flavors. The trapping scheme takes into 
account the optical 
depth at any location inside the star to reduce the neutrino release when the
diffusion timescale becomes long. Since the emission of electron neutrinos and
antineutrinos can change the electron lepton number of the stellar medium, we also
solve a continuity equation for this quantity. A detailed discussion of the 
equations and a more complete explanation of the numerical methods can be found in
Ruffert et al.~(1996, 1997a).

The numerical code and the input physics described in 
Ruffert et al.~(1996, 1997a) 
have been changed and improved in a number of aspects. 

\smallskip\noindent
{\em (i) Grid:}\\
The cartesian grid for the 
simulations presented there had a side length of only 82$\,$km. Therefore a 
significant amount of mass was swept off the grid although it would not have become
unbound. For this reason we enlarged the computational volume to 328--656$\,$km
by introducing up to five levels of nested grids (Fig.~\ref{fig:1}). 
This ensures good numerical
resolution near the grid center so that the two neutron stars are represented
in the new calculations with a similar accuracy as the best resolved Model~A128 in 
Ruffert et al.~(1996, 1997a). The larger computational volume allows us to trace
the matter with high angular momentum, which is flung out to large distances but
returns and is added to the cloud of gas that surrounds the compact merger remnant.

\smallskip\noindent
{\em (ii) EoS table:}\\
Another problem in the previous calculations was that
under certain extreme conditions (e.g., extreme heating as in case of Model~C64)
the boundaries of the EoS table were hit. Therefore we expanded the temperature
range in a new EoS table to cover values from 10$^{-2}\,$MeV to 100$\,$MeV,
and reduced the minimum density to $5\times 10^7\,$g$\,$cm$^{-3}$. 

\smallskip\noindent
{\em (iii) Environmental density:}\\
Since the simulations are done in an Eulerian reference frame, the density
outside the neutron stars cannot be set to zero but only to a small value, i.e.,
a value small compared to the average density in the stellar interior. In order
not to influence the dynamical behavior of the ejected stellar fragments by gas 
that is swept up in the
surroundings, the density of this ambient medium should be chosen as low as possible.
Therefore we used a value of $10^8\,$g$\,$cm$^{-3}$ 
in the new calculations instead of  
$10^9\,$g$\,$cm$^{-3}$ previously. The total mass of this dilute gas on the grid
is therefore significantly less than $10^{-3}\,$M$_{\odot}$. Since only an 
extremely small fraction of this gas interacts with the neutron star matter,
the dynamical effect of this medium is negligible.

\smallskip\noindent
{\em (iv) Temperature determination and entropy equation:}\\
The employed hydrodynamics code solves the energy equation for the total
specific energy, which is the sum of the specific internal and kinetic energies.
Compared to integrating an equation for the internal energy alone, this
has the advantage that without gravity the energy equation is written in
a fully conservative way. The effects of gravity are included by source terms.
It has, however, the disadvantage that the internal energy, from which the 
temperature is determined, has to be calculated as the difference of total and kinetic
energies. If both the latter energies are large, a small value of the internal energy
can be significantly affected by numerical noise, and the temperature determination 
can become inaccurate. This is usually not a problem, but does become problematic 
in the case of extremely
degenerate neutron star matter, which has a very small heat capacity. 
Only minor variations of the internal energy can then lead to much larger 
changes of the temperature.
For these reasons the temperature is a 
quantity which is very sensitive to numerical deficiencies. As a consequence,
it was not possible to start the simulations in Ruffert et al.~(1996, 1997a) 
with cold neutron stars. A stable temperature evolution could be obtained when the
thermal energy density in the initial configuration was assumed to be two per 
cent of the degeneracy energy density. With this prescription, the central temperature
of the neutron stars was several MeV and the surface temperature about half an 
MeV initially. Nevertheless, the corresponding neutrino luminosities were 
negligibly small, and the total thermal energy was so tiny compared to the 
other energies (internal, gravitational and kinetic) that its effect on the 
dynamics was unimportant.

In order to avoid these problems in the temperature determination and to have the
freedom of starting with colder neutron stars, we decided to use the entropy 
instead of the energy density for temperature calculations. The entropy evolution
was followed by a separate entropy equation which was integrated in addition to
the hydrodynamic conservation laws. In an Eulerian frame of reference, the
corresponding continuity equation is
(making use of Einstein's summation convention)
\begin{equation}
{\partial\over \partial t}\rund{s\,n_{\mathrm b}} + 
{\partial(s\,n_{\mathrm b} v^j)\over \partial x^j}
\,=\,S_{\nu} + S_{\mathrm {sh}} + S_{\mathrm {vis}} \, .
\label{eq:1}
\end{equation}
Here $s$ means the matter entropy per nucleon, $n_{\mathrm b} = \rho/u$ the
baryon number density ($u$ being the atomic mass unit, $\rho$ the rest mass density), 
and $v^j$ the cartesian
components of the velocity vector. The divergence of the entropy flux was written
in its cartesian form (summation is applied when an index appears twice), and the 
source terms on the right hand side are the rates of change of the entropy density
due to neutrino production, shock dissipation, and shear and bulk viscosity 
effects, respectively. 

The entropy generation rate per unit volume by neutrino processes is 
(e.g., Cooperstein 1988)
\begin{equation}
S_{\nu}\,=\,{S_{\mathrm E}\over k_{\mathrm B}T} - S_{\mathrm L}\rund{
\psi_e + \psi_p - \psi_n}\, ,
\label{eq:2}
\end{equation}
where $S_{\mathrm E}$ and $S_{\mathrm L}$ are the effective energy loss 
rate and the effective
lepton number source term as defined in Appendix~B of Ruffert et al.~(1996),
$k_{\mathrm B}T$ is the temperature in MeV and the $\psi$'s denote the
degeneracy parameters (chemical potentials divided by the temperature) of
electrons, protons and neutrons (including the rest mass energies). 
The neutrino source terms as given by Ruffert et al.~(1996) are only 
evaluated for temperatures above about 0.5$\,$MeV. Below this threshold they 
are switched off, because the assumptions employed in their calculation are 
not valid any more.

The entropy generation rate per unit volume by shocks is given according 
to the tensor formalism of Tscharnuter \& Winkler (1979) as
\begin{equation}
S_{\mathrm {sh}}\,=\,-\,{Q_k^i \epsilon_i^k\over k_{\mathrm B}T} \,,
\label{eq:3}
\end{equation}
with the mixed tensor $Q_l^m$ of the viscous pressure given by 
\begin{equation}
Q_l^m \,=\, \cases {l^2\rho\,{\displaystyle {\partial v^k\over \partial x^k}}
\rund{\epsilon_l^m - {\displaystyle {\delta_l^m\over 3}}\,
{\displaystyle {\partial v^k\over \partial x^k}}}\,,
&if\ \ \ ${\displaystyle {\partial v^k\over \partial x^k}}<0$\, ;\cr
0\,,\phantom{\int\limits^0_0} &otherwise\, , \cr}
\label{eq:4}
\end{equation}
and the mixed tensor $\epsilon_l^m$ of the symmetrized gradient of the velocity
field defined by
\begin{equation}
\epsilon_l^m \,=\, {1\over 2}\,\rund{{\partial v^m\over \partial x_l} +
{\partial v_l\over \partial x^m}}\, ,
\label{eq:5}
\end{equation}
where Einstein's summation convention is used ($\partial v^k/\partial x^k$
therefore means the divergence of the velocity vector in cartesian coordinates), 
and $\delta_l^m$ is the mixed unity tensor. The characteristic length $l$ 
is of the order of the local width of the grid, $l = f\cdot \Delta x$.
We calibrated the proportionality factor $f$ such that the entropy jump
across a shock as calculated with our hydrodynamics code is reproduced by the
entropy generation according to Eqs.~(\ref{eq:3})--(\ref{eq:5}). 
We found best agreement for the choice of $f = 1.8$. 

The entropy generation rate per unit volume due to shear and bulk viscosity 
can be written (using again the summation convention)
as (e.g., Shapiro \& Teukolsky 1983, Landau \& Lifschitz 1991)
\begin{eqnarray}
S_{\mathrm {vis}}\,=\, {1\over k_{\mathrm B}T}\,\Biggl[ {1\over 2}\,\eta\,
\rund{{\partial v^i\over \partial x^j} + {\partial v^j\over \partial x^i}}^{\! 2}
&-& {2\over 3}\,\eta\,\rund{{\partial v^k\over \partial x^k}}^{\! 2}
\Biggr.
\nonumber \\
&+& \Biggl. \zeta\,\rund{{\partial v^k\over \partial x^k}}^{\! 2}\, \Biggr] \, ,
\label{eq:6}
\end{eqnarray}
where $\eta$ is the dynamic shear viscosity coefficient and $\zeta$ the bulk 
viscosity coefficient. Solving the Euler equations for an ideal fluid, shear 
viscosity effects are only caused by the numerical viscosity of our hydrodynamics
code, which we describe by the ansatz $\eta = \alpha\rho v \Delta x$.
With a typical grid resolution $\Delta x$ between $10^4\,$cm and $10^5\,$cm 
one empirically finds values for $\alpha$ between $5\times 10^{-4}$ and 
$5\times 10^{-3}$ (Janka et al.~1999). We used a representative number
of $\alpha = 2\times 10^{-3}$ in the simulations discussed below.

In hot neutron star matter with the proton fraction exceeding a critical 
lower limit, bulk  viscosity can be strongly enhanced by the direct URCA 
processes of electron neutrino and antineutrino production and absorption
(Haensel \& Schaeffer 1992).
For matter composed of neutrons ($n$), protons ($p$) and electrons ($e$)
with trapped neutrinos (but with no trapped lepton-number
excess, i.e., if $\psi_e +\psi_p -\psi_n \ll 1$) one can write an 
approximate expression for the bulk viscosity 
coefficient as $\zeta \sim 10^{24}(Y_p\rho/\rho_0)^{1/3}\,$g/(cm$\,$s)
with $Y_p = Y_e$ being the proton fraction and $\rho_0\approx 2.5\times 10^{14}\,$
g/cm$^3$ the density of normal nuclear matter (Haensel \& Schaeffer 1992,
Sawyer 1980, van den Horn \& van Weert 1981). Although the numerical factor was 
taken somewhat larger than estimated by Haensel \& Schaeffer (1992),
we found the entropy generation rate associated with the  
bulk viscosity term to be negligibly small. 

The entropy loss rate due to
neutrino emission, expressed by the source term $S_{\nu}$, is tiny initially,
but becomes (globally, i.e.\ as an integral over all grid cells)
comparable to the (positive) shear viscosity term $S_{\mathrm {vis}}$ 
after several milliseconds, when the neutron stars have merged to
a hot, rapidly and differentially spinning object. Earlier than this,
in particular prior to the merging, the time is too short for shear viscosity
to raise the entropy, and the temperature is too low for neutrinos to make
any effect. When the neutron stars begin to touch (roughly half a millisecond
after the simulations were started), the shock dissipation term 
$S_{\mathrm {sh}}$ becomes clearly the dominant one globally,
about twice to twenty times bigger than $S_{\mathrm {vis}}$.

Using Eq.~(\ref{eq:1}) for evolving the entropy, an updated value of the
temperature is obtained in a predictor-\-cor\-rec\-tor step which ensures
second order accuracy for the time integration. In a first step
one evaluates the entropy source terms with the old 
temperature, then solves Eq.~(\ref{eq:1}) to get an estimate for the new 
entropy and thus for the new temperature, and then solves 
Eq.~(\ref{eq:1}) a second time with source terms computed with an average
value of the old and estimated new temperature.

The temperature thus obtained from the entropy equation is used to 
calculate the neutrino source terms in the hydrodynamic conservation
laws of mass, momentum, energy, and lepton number, which still describe 
the evolution of the stellar fluid. This is not fully consistent
and should therefore not be considered as the necessarily better 
treatment. Instead, it is meant as an alternative approach
which allows one to test the uncertainties associated with the 
temperature determination and the corresponding effects of neutrinos.

\smallskip\noindent
{\em (v) Neutrino treatment:}\\
Let us conclude this section with a few remarks about the treatment of neutrinos.
The energy radiated in neutrinos during the computed evolution (about 10$\,$ms)
is typically more than an order of magnitude smaller than the total energy 
emitted in gravitational waves. Whereas neutrino emission is very small during the
first five milliseconds, it increases later and dominates the energy loss 
in the second half of the computed evolution. For the dynamical phase of 
the merging process (which lasts only a few milliseconds after the start of the 
simulations), neutrino source terms are therefore insignificant.
Of course, the treatment of neutrino effects by using a trapping scheme
(Ruffert et al.~1996, 1997a) is a strong simplification of the true problem.
An exact treatment would require three-dimensional, time-dependent 
(general relativistic) transport of neutrinos and antineutrinos
in a moving neutron star medium, which is extremely optically thick in 
dense and hot regions and transparent near the stellar surface and
in regions where the matter is cold with temperatures below about 1$\,$MeV. 
Solving this problem is currently not feasible, but we are convinced that our 
trapping scheme is a good first approach which yields an approximate
description of effects connected with the production of 
neutrinos, and a reasonably good estimate for the total luminosity of neutrinos.

Neutrinos stream off freely as soon as they are produced in transparent
matter, thus changing the quantity $X_{\mathrm m}$ (i.e., energy, entropy 
or lepton number) of the stellar medium 
according to the rate given by the neutrino source term $S_{\nu}$:
\begin{equation}
{\partial X_{\mathrm m}\over \partial t}\,=\, S_{\nu} \, .
\label{eq:7}
\end {equation}
In contrast, the dominant mode of energy, entropy and lepton number loss is by
neutrino diffusion when neutrinos are in equilibrium with the stellar medium
at high optical depths. This is expressed by the equation
\begin{equation}
{\partial (X_{\mathrm m} + X_{\nu})\over \partial t}\,=\, - \nabla F_{\nu} \, .
\label{eq:8}
\end {equation}
The description used in the trapping scheme is guided by these cases.
It calculates the loss terms for energy (and entropy) and
lepton number from the local neutrino emission rates in the optically thin
regime, and from the rate of diffusive depletion of the equilibrium
densities of neutrino number and energy in the optically thick regime. The
transition between both limits is done with a smooth interpolation based on
the local diffusion timescale, which is estimated from the optical
depth to the stellar surface (see Appendix~B in Ruffert et al.~1996).
In the equilibrium regime, however, we neglect the contributions of neutrinos
to the internal energy density and to the entropy density, or, in other words,
we neglect that neutrinos in equilibrium contribute to the heat capacity of 
the stellar matter (an effect which would affect the computed temperature). A
similar approximation is also applied for the (electron) lepton number because
we solve a continuity equation only for the net electron number density
(number density of electrons minus number density of positrons)
and neglect the lepton numbers carried by electron neutrinos and 
antineutrinos. These approximations can be justified by the fact that 
the chemical potential of the electrons is much larger than that of neutrinos.
Therefore the net lepton number of electron neutrinos minus antineutrinos is
very small, and the summed energy densities of neutrinos
and antineutrinos of all flavors are always less than $\sim\,$10\% of the energy 
density of electrons (plus positrons, if present) plus photons, and even smaller
compared to the total internal energy density of the gas, which also includes the
contributions of nucleons and nuclei. For the same reason, we also made no effort
to take into account the neutrino pressure (or momentum transfer to the stellar
medium when neutrinos start to decouple).

The use of a trapping scheme means that neutrino diffusion
or propagation relative to the stellar medium and neutrino advection along with
the fluid motion are ignored. Therefore the corresponding
energy and lepton number (and entropy) transport inside the star are disregarded. 
These effects are likely not to be very important within the few milliseconds
of the computed evolution, because the diffusion timescale in the dense, hot
interior of neutron stars is several seconds and neutrino lepton number and
energy are small compared to the medium quantities. 
Even in the less dense region between
neutrinosphere and nuclear core the changes of energy and lepton number due
to local processes (which are evaluated with the trapping scheme) should
dominate the transfer of energy and leptons by neutrinos which stream from one 
region of the star to another. Therefore we think that the trapping scheme
is suitable to account for the main effects which play a role on millisecond
timescales. If the simulations were continued for a longer time, tens or hundreds
of milliseconds, the transport effects would, of course, have to be included.

The trapping scheme fails to yield a suitable approximation when and where neutrino 
heating is stronger than neutrino cooling. 
This is the case outside of the neut\-ri\-no\-sphere.
Here high-energy neutrinos transfer energy to the cooler stellar gas via scattering
and absorption reactions. This energy deposition causes an outflow of baryonic mass,
the so-called neutrino-driven wind, which has been investigated in some detail 
for cooling proto-neutron stars in type-II supernovae (Duncan et al.~1986,
Woosley 1993b, Qian \& Woosley 1996). 
It is a major disadvantage that the trapping scheme
does not allow one to study this interesting phenomenon in the context of 
merging neutron stars. We shall try to remove this deficiency in future 
improvements of our code.

\begin{table*}[t]
\caption[]{
Computed models with their characterizing parameters and results.
The models differ concerning the chosen initial spins of the 
neutron stars. In several cases the temperature evolution was followed
by using an entropy equation, starting from ``high''
(``ent~wa'') or low initial temperature (``ent~co''). Also a special 
choice of the grid is indicated by the model name and in the column with
remarks.
$N$ is the number of grid zones per dimension in the orbital plane, 
$g$ the number of levels of the nested grid,
$L$ the size of the largest grid, 
$l$ the size of the smallest zone,
$M_1$ and $M_2$ are the masses of the two neutron stars,
$a_0$ is their initial center-to-center distance,
$k_{\rm B}T_0$ the initial maximum temperature (reached at the center of the neutron stars), 
$t_{\rm sim}$ the computed period of time of the evolution,
$M_{\rho<11}$ is the gas mass with a density below $10^{11}\,$g/cm$^3$ at the end of the simulation,
$M_{\rm d}$ the gas mass with specific angular momentum larger than the Keplerian angular 
momentum at a radius equal to three Schwarzschild radii of the merger remnant,
$M_{\rm g}$ the mass of the gas which leaves the grid,
$M_{\rm u}$ the gas mass which can become unbound, and
$k_{\rm B}T_{\rm mx}$ the maximum temperature reached on the grid during the simulation.
The $>$ signs indicate that the numbers are still changing when the simulations were stopped. 
}
\begin{flushleft}
\tabcolsep=1.8mm
\begin{tabular}{lllccccccccccccccc}
\hline\\[-3mm]
model & remark & spin & $N$ & $g$ & $L$ & $l$ & $M_1$ & $M_2$ & $a_0$ & 
   $k_{\rm B}T_0$ & 
   $t_{\rm sim}$ &  $M_{\rho<11}$ & 
   $M_{\rm d}$ & $M_{\rm g}$ & $M_{\rm u}$ & 
   $k_{\rm B}T_{\rm mx}$ \\
 & & & & & km & km & M$_\odot$ & M$_\odot$ & km & 
   {\scriptsize MeV} &
   ms & {\scriptsize M$_\odot$/100} & 
   {\scriptsize M$_\odot$/100} & {\scriptsize M$_\odot$/100} &
   {\scriptsize M$_\odot$/100} & 
   {\scriptsize MeV} \\[0.3ex] \hline\\[-3mm]
A32 &   ---    & none & ~32 & 4 & 328 & 1.28 & 1.6 & 1.6 & 42. & 5.44 & 10.
    & 4.5 & 6.1 & 0.78 & 0.03 & 53. \\
A64 &   ---    & none & ~64 & 4 & 328 & 0.64 & 1.6 & 1.6 & 42. & 5.47 & 10.
    & 5.8 & 8.0 & 1.78 & 0.23 & 39. \\[0.7ex]
B32 &   ---    & solid& ~32 & 4 & 328 & 1.28 & 1.6 & 1.6 & 42. & 6.76 & 10.
    & 6.9 & 21. & 8.3  & 1.6  & 39.  \\
B32w& ent wa   & solid&~32 & 4 & 328 & 1.28 & 1.6 & 1.6 & 42. & 5.44 & 10.
    & 6.5 & 19. & 9.6 & 2.2   & 30.   \\ %Grbp
B32w'& ent'wa  & solid&~32 & 4 & 328 & 1.28 & 1.6 & 1.6 & 42. & 5.44 & 10.
    & 6.1 & --- & 9.7 & 2.2   & 49.    \\ %Grbt
B32c& ent co   & solid&~32 & 4 & 328 & 1.28 & 1.6 & 1.6 & 42. & 0.05 & 10.
    & 6.5 & 19. & 9.6 & 2.1   & 29.   \\  %Grbf
B32$^5$c & ent co  & solid&~32 & 5 & 656 & 1.28 & 1.6 & 1.6 & 42. & 0.05 & 24.
    & 14. & 18. & 4.5 & 1.9   & 30.   \\  %Grbl
B64 &   ---    & solid& ~64 & 4 & 328 & 0.64 & 1.6 & 1.6 & 42. & 6.25 & 10.
    & 5.7 & 25. & 9.2  & 2.4  & 39.  \\
B64c & ent co  & solid&~64 & 4 & 328 & 0.64 & 1.6 & 1.6 & 42. & 0.05 & 10.
    & 5.5 & 25. & 9.1 & 2.4   & 40.   \\[0.7ex] %GrbF
C32c & ent co  & anti &~32 & 4 & 328 & 1.28 & 1.6 & 1.6 & 42. & 0.05 & 10.
    & 4.2 & 1.7 & 0.13 & 0.005 & 58. \\  %Grbg
C64c & ent co  & anti &~64 & 4 & 328 & 0.64 & 1.6 & 1.6 & 42. & 0.05 & 10.
    & 5.7 & 6.0 & 0.35 & 0.0085 & 69. \\  %GrbG
C128c& ent co  & anti &128 & 4 & 328 & 0.32 & 1.6 & 1.6 & 42. & 0.05 & 2.
    & $>$0.3 & $>$1. & $>$0. & $>$0.  & 78.   \\[0.7ex]  %GrbH
O32 &   ---  & oppo & ~32 & 4 & 328 & 1.28 & 1.6 & 1.6 & 42. & 0.05 & 10.
    & 4.0 & 9.0 & 1.57 & 0.36 & 66. \\
O64 &   ---  & oppo & ~64 & 4 & 328 & 0.64 & 1.6 & 1.6 & 42. & 0.05 & 10.
    & 3.4 & 8.4 & 1.19 & 0.19 & 89. \\[0.7ex]
S32 &   ---   & solid& ~32 & 4 & 328 & 1.28 & 1.2 & 1.2 & 42. & 4.68 & 11.6
    & 6.0 & 16. & 7.9  & 2.4  & 32.  \\
S64 &   ---    & solid& ~64 & 4 & 328 & 0.64 & 1.2 & 1.2 & 42. & 4.71 & 10.
    & 6.5 & 23. & 7.5  & 2.0  & 35.  \\[0.7ex]
D32 &   ---    & solid& ~32 & 4 & 400 & 1.56 & 1.8 & 1.2 & 46. & 7.16 & 10.
    & 7.1 & 14. & 8.8  & 2.7  & 39.  \\
D64 &   ---    & solid& ~64 & 4 & 400 & 0.78 & 1.8 & 1.2 & 46. & 7.19 & 13.
    & 7.0 & 13. & 9.3  & 3.8  & 35.  \\
D64$^3$ & 3 grids  & solid& ~64 & 3 & 400 & 1.56 & 1.8 & 1.2 & 46. & 7.16 & 5.
    & $>$2.7& 22. & $>$0.8 & $>$0.4 & $>$33. \\    %GrbE
D128$^3$& 3 grids  & solid& 128 & 3 & 400 & 0.78 & 1.8 & 1.2 & 46. & 7.19 & 2.
    & $>$0.2& $>$5. & $>$0.  & $>$0.  & $>$12. \\[0.7ex]  %Grbe
\hline
\end{tabular}
\end{flushleft}
\label{tab:models1}
\end{table*}

\section{New simulations}
\label{sec:simulations}

In this section we shall describe the different models and will review 
the most important results of our simulations. 

Binaries with different baryonic masses of the neutron stars, 1.2, 1.6, and 
1.8$\,$M$_{\odot}$,
were considered, with different mass ratios (1:1 and 1:1.5), and with
different neutron star spins, where we distinguish the four cases of initially
irrotational systems (``none''), synchronous 
(or `tidally locked') 
rotation (``solid''),
counterrotation (``anti''), and opposite spin directions (``oppo'') of the 
two neutron stars. The angular velocity due to the spins, which is added to the 
orbital motion, is equal to the angular velocity of the orbit at the chosen initial
distance of the two stars, but the spin directions are varied between the
four cases (see also Ruffert et al.~1996). In addition, we computed models with 
different grids, i.e., with different zone sizes of the finest grid (and thus 
different numerical resolution) as well as different numbers of grid levels.
Finally, we   
used the entropy equation to follow the temperature history of our models,
and started the computations with different temperatures of the neutron stars,
``warm'' models with a central temperature of 5--7$\,$MeV, satisfying the
requirement that the thermal energy at the beginning is two per cent of the 
internal energy, and ``cold'' models with a constant initial temperature of only 
0.05$\,$MeV.

\subsection{Initial conditions}
\label{sec:inicon}

The initial configuration consists of two spherically symmetric neutrons
stars, constructed as hydrostatic equilibrium solutions for
Newtonian gravity and the EoS of Lattimer \& Swesty (1991). The temperature
of cold and warm models is chosen as mentioned above, and the profile of the 
electron 
fraction, $Y_e(r)$, corresponds to the equilibrium state of cold, deleptonized,
neutrino-transparent neutron stars (i.e., the chemical potential of electron 
neutrinos vanishes throughout the star). The 1.6$\,$M$_{\odot}$ neutron star
has a radius of about 15$\,$km, the 1.2$\,$M$_{\odot}$ star is slightly 
smaller and the 1.8$\,$M$_{\odot}$ star a little bigger. This scaling
of mass and radius indicates that the effective adiabatic index of the stellar
matter is larger than two (see Shapiro \& Teukolsky 1983). The initial 
center-to-center distance $a_0$ of the two stars was set to a value such that the
stars could finish about one full revolution before the final plunge occurred
at the radius of tidal instability. Circular orbits were assumed for all models.
The initial orbital velocity was chosen according to the inspiraling motion 
of two point masses $M_1$ and $M_2$ at separation $a_0$ in response to their
emission of gravitational waves. From the quadrupole formula the
angular velocity and the radial velocity can be calculated as 
(Cutler \& Flanagan 1994):
\begin{eqnarray}
\omega &=& \sqrt{{G(M_1 + M_2)\over a_0^3}} \, , 
\label{eq:9}\\
v_r    &=& \dot a   \,=\, -\,{64\over 5}\,{G^3\over c^5}\,{1\over a_0^3}\,
             M_1M_2(M_1+M_2) \, .
\label{eq:10}
\end{eqnarray}
More details about the treatment of the initial conditions can be found in
Ruffert et al.~(1996).

\subsection{Models}
\label{sec:models}

Table~\ref{tab:models1} contains a list of computed models with their specific
characteristics. Basically we distinguish models of types A,B,C,O,S and D.
Models A,B,C and O are our standard cases with two equal neutron stars with
baryonic masses of 1.6$\,$M$_{\odot}$ and initially irrotational (spin: ``none''), 
corotational (``solid''), counterrotational (``anti'')
and opposite directions (``oppo''), respectively, for the spins in 
the initial state. The S-models are computed with smaller,
1.2$\,$M$_{\odot}$ neutron stars, and the D-models with two different neutron 
stars of 1.2$\,$M$_{\odot}$ and 1.8$\,$M$_{\odot}$. (Note that Model~V64 in 
Janka et al.~1999 is identical to Model~C64c of this work.)

The names of the models carry information also
about the number of zones of each grid level (32, 64, or 128),
and about the number of grid levels, indicated by a superscript, if it is not 
the standard value of 4. Since we take the extension of the grid in the
$z$-direction perpendicular to the orbital plane to be only half as big
as in the orbital plane and, in addition, assume equatorial symmetry, the 
grids have 32$\times$32$\times$8 or 64$\times$64$\times$16 or
128$\times$128$\times$32 zones, respectively. A higher grid level has only
half the resolution (twice the zone size) of the level below. 
With an equatorial length and width of the computational volume between 
328$\,$ and 656$\,$km, the smallest zones have a side length between 
0.32$\,$km and 1.56$\,$km.
For fixed computational volume a smaller number of grid levels implies that
the individual grids are bigger, and avoids that the two neutron stars
cross the boundary of the innermost grid during spiral in, as they do for
our standard case of four levels of nested grids (see Fig.~\ref{fig:1}).
Reducing the number of grid levels therefore allows one to test the
corresponding numerical effects.

Use of the entropy instead of the energy to calculate the temperature
(which is then used for computing the neutrino emission)
is indicated by an extension of the model name. The letter ``w'' marks
the ``warm'' models, ``c'' the ``cold'' ones. Since our stars have to be 
embedded by a medium with finite density on the Eulerian grid, the motion
of the stars through this medium creates a shock wave at the stellar surfaces,
where the temperature increases in a narrow ring of grid zones. This 
effect is energetically umimportant, and also does not lead to a 
significant increase of the neutrino luminosity as long as the 
volume and the density of the heated medium stay low. However it is 
unphysical and can lead to an inflation of the surface layers of the neutron 
stars. Therefore we try to reduce it by localizing grid cells 
with temperature spikes in the shocked region
below a certain density
and resetting the temperature
there to an average value of 
the surrounding grid zones. The influence of this manipulation has to
be tested. On the one hand we did this by computing models with increasingly
better resolution (where the disturbing effects occured in a smaller volume
of space), on the other hand we changed the procedure for detecting the 
zones with unphysical temperatures. A corresponding test model is B32w'.

%********************************** fig.2 ******************************************
\begin{figure*}[htp!]
 \vspace{25cm}
\caption[]{
Density (left) and temperature (right) distribution in the orbital
plane of Model~A64 for different times after the start of the simulation.
The arrows in the density plots indicate the velocity field. The density
is given in g$\,$cm$^{-3}$ with
contours spaced logarithmically in steps of 0.5~dex. The temperature
is measured in MeV, its contours are labeled with the corresponding
values, the bold contours being 10~MeV, 20~MeV, etc.
Note the different scales of the axes of the plots for different times
\label{fig:2}
}
\end{figure*}
%********************************** fig.2 ******************************************

%********************************** fig.3 ******************************************
\begin{figure*}[htp!]
 \vspace{25cm}
\caption[]{
Same as Fig.~\ref{fig:2}, but for Model~B64
\label{fig:3}
}
\end{figure*}
%********************************** fig.3 ******************************************

%********************************** fig.4 ******************************************
\begin{figure*}[htp!]
\vspace{25cm}
\caption[]{
Same as Fig.~\ref{fig:2}, but for Model~C64c
\label{fig:4}
}
\end{figure*}
%********************************** fig.4 ******************************************

%********************************** fig.5 ******************************************
\begin{figure*}[htp!]
\vspace{25cm}
\caption[]{
Same as Fig.~\ref{fig:2}, but for Model~O64
\label{fig:5}
}
\end{figure*}
%********************************** fig.5 ******************************************

%********************************** fig.6 ******************************************
\begin{figure*}[htp!]
\vspace{25cm}
\caption[]{
Same as Fig.~\ref{fig:2}, but for Model~D64
\label{fig:6}
}
\end{figure*}
%********************************** fig.6 ******************************************

%********************************** fig.7 ******************************************
\begin{figure*}[htp!]
\vspace{25cm}
\caption[]{
Density (left) and temperature (right) in the x-z-plane and 
y-z-plane perpendicular to the orbital plane at the end of the
computed evolution of Models~B64, D64, and O64. The plots correspond
to the last moments shown in Figs.~\ref{fig:2}, \ref{fig:3}, and
\ref{fig:4}, respectively
\label{fig:7}
}
\end{figure*}
%********************************** fig.7 ******************************************

\section{Results}
\label{sec:results}

In this section the results of our new simulations will be described 
as far as important differences to our previously published models
showed up, or interesting effects occurred in dependence of the 
parameters varied between the models.

\subsection{Dynamical evolution and gravitational-wave emission}

Figures~\ref{fig:2}--\ref{fig:7} show the hydrodynamical results for five
of the models listed in Table~\ref{tab:models1}. The displayed cases 
include Model~A64 with two equal, initially nonrotating neutron stars
(Fig.~\ref{fig:2}), Model~B64 with two equal, initially corotating stars
(Fig.~\ref{fig:3}), Model~C64 with two equal, initially counterrotating 
stars (Fig.~\ref{fig:4}), Model~O64 with equal neutron stars with the spins 
pointing in opposite directions (Fig.~\ref{fig:5}), and Model~D64 with 
neutron stars of different masses in initially locked rotation 
(Fig.~\ref{fig:6}). Figure~\ref{fig:7} provides cuts perpendicular to
the orbital planes for Models~B64, O64, and D64, whereas the other plots
display the density and temperature distributions in the orbital plane. 

Compared to our previously published simulations (Ruffert et al.~1996, 1997a), 
Models~A64, B64, and C64 were computed with a finer resolution around the
grid center and simultaneously a significantly larger volume
by employing nested grids. Corresponding models with 32 zones on the finest
grid level have a similar resolution as the standard models in our preceding
papers. In addition, Model~C64 was started with
a lower initial temperature and the temperature history was followed by using
the entropy equation. 

The displayed results reveal a significant dependence
of the dynamical evolution on the neutron star masses and spins. In the 
corotating case (B64) pronounced spiral arms form, which are much less 
strongly developed in the A and C models and are present only for the corotating
component in Model~O64. In case of the different neutron star masses (D64) the
less massive star is stretched to a long, banana shaped object before it is 
finally wrapped around the more massive component. Prior to this, a sizable
amount of matter is flung out from the outward pointing end of the smaller
star. This material forms a spiral arm which expands away from the merger site.

\begin{figure}
  \epsfxsize=9.2cm \epsfclipon \epsffile{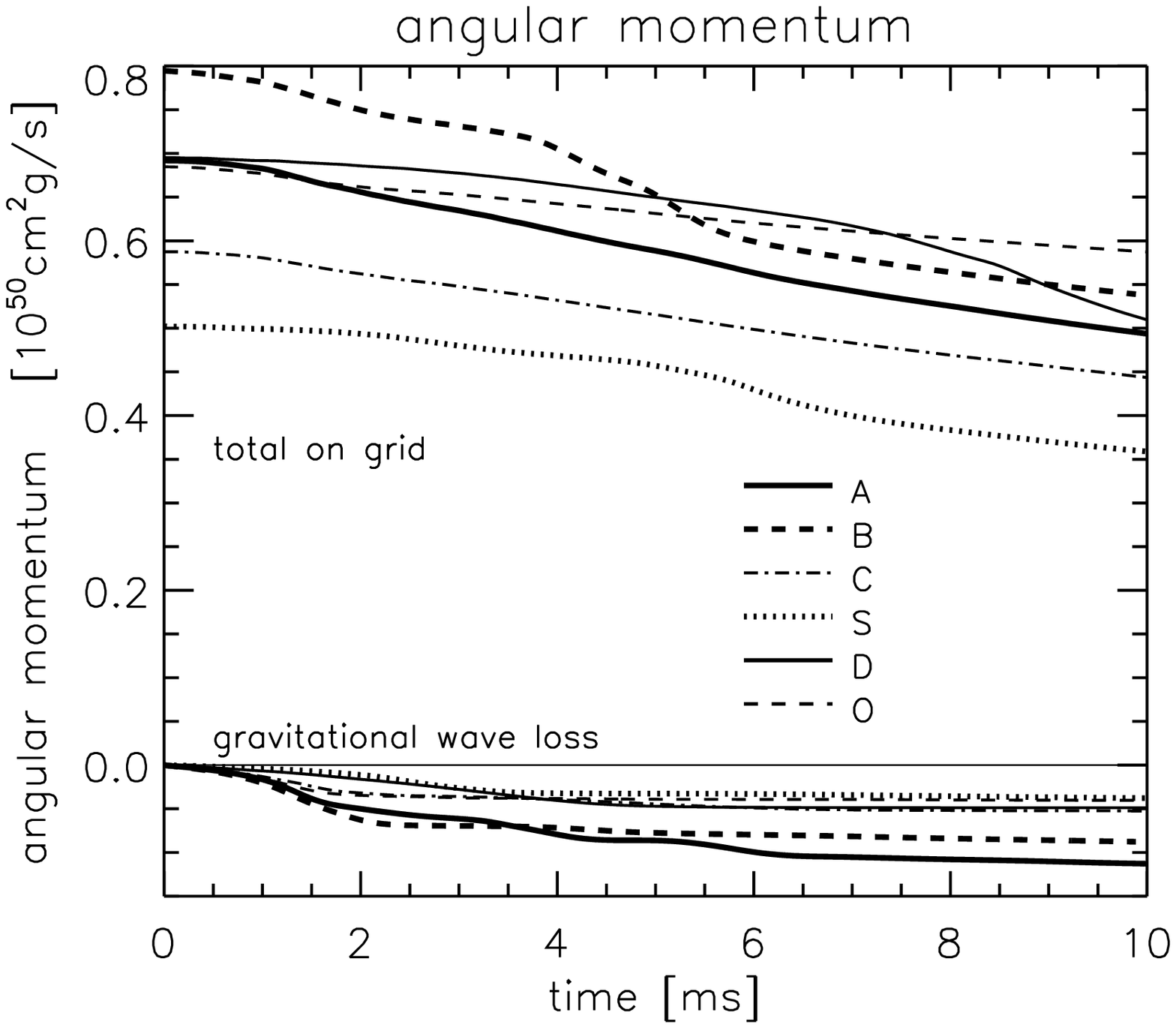}
  \caption[]{Total angular momentum (component perpendicular to the 
   orbital plane) of the matter
   on the grid (upper curves) and cumulative angular momentum loss
   by gravitational wave emission (lower curves) as functions of
   time for different models}
\label{fig:8a}
\end{figure}

The initial total angular momentum of the binary systems varies with 
the spins of the two components. Among the models with the same masses 
of the two neutron stars, the total angular momentum is largest for
the B cases (neutron stars corotating with the orbit), smaller for the 
A (no neutron star spins) and O models (spins in opposite directions), 
and smallest for the C models (both neutron stars in counterrotation with 
the orbital motion) (Fig.~\ref{fig:8a}). During the computed evolution,
about 10\% of the angular momentum are removed by the emission of 
gravitational waves (Fig.~\ref{fig:8a}). This
leads to correlations between maxima of the gravitational-wave
luminosity and periods of a decrease of the angular momentum on the grid
(compare Figs.~\ref{fig:8a} and \ref{fig:9a}). More angular momentum
is carried away by the matter that leaves the boundaries of the 
computational grid. While the gravitational-wave emission peaks within
the first 2 milliseconds, the effect due to the mass loss becomes important 
only later than 4--6~ms after the start of the simulations.
In addition, the code does not conserve the angular momentum exactly. The
violation depends mainly on the resolution of the finest grid level, where
the bulk of the matter is located. For the models with a cell size of
0.64~km on the central grid, less than 10\% of the initial angular momentum
were destroyed within about 10~ms of evolution. This, unfortunately, does 
not reach the excellent quality of the conservation of energy by our code
(Fig.~\ref{fig:8b}).

The total angular momentum of the coalescing binary system determines
the structure and the dynamical state of the very compact central body 
of the merger remnant, which contains the bulk ($\ga 90$\%) of the system 
mass. The wobbling and ringing of this body after the merging is sensitive to
the fluid motions and thus to the value and distribution
of angular momentum in its interior. The plots in 
Fig.~\ref{fig:8} show that the two stars in Model~B64 fuse smoothly and
form a centrally condensed body immediately after the plunge, whereas
in Model~A64, and even stronger in Model~C64, two density maxima can still
be distinguished later. In particular in the counterrotating Model~C64
the distance of these density maxima varies with time, indicating that
the remnant remains in a perturbed internal state. Correspondingly, the
gravitational waveform of Model~C64 (Fig.~\ref{fig:9}) reveals a 
low-freqency modulation on top of the high-frequency structure which is
caused by the oscillation period of the compact remnant. This feature is
essentially absent in Model~B64. 

%********************************** fig.8a *****************************************
%********************************** fig.8a *****************************************

%********************************** fig.8b ******************************************
\begin{figure*}[htp!]
 \vspace{7cm}
\caption[]{
Total internal, kinetic, and gravitational potential energies of the matter
on the grid as functions of time for Models~B64 and C64c. The total energy
is calculated by adding up these energies plus the energy that was
emitted in gravitational waves. It should be conserved, because the gas that
leaves the grid, has a total energy very close to zero
\label{fig:8b}
}
\end{figure*}
%********************************** fig.8b ******************************************

%********************************** fig.8 ******************************************
\begin{figure*}[htp!]
\vspace{10.5cm}
\caption[]{
The separation of the density maxima and the values of the 
maximum density on the grid as functions of time for different models
\label{fig:8}
}
\end{figure*}
%********************************** fig.8 ******************************************

%********************************** fig.9a ******************************************
\begin{figure*}[htp!]
\begin{tabular}{cc}
  \epsfxsize=8.8cm \epsfclipon \epsffile{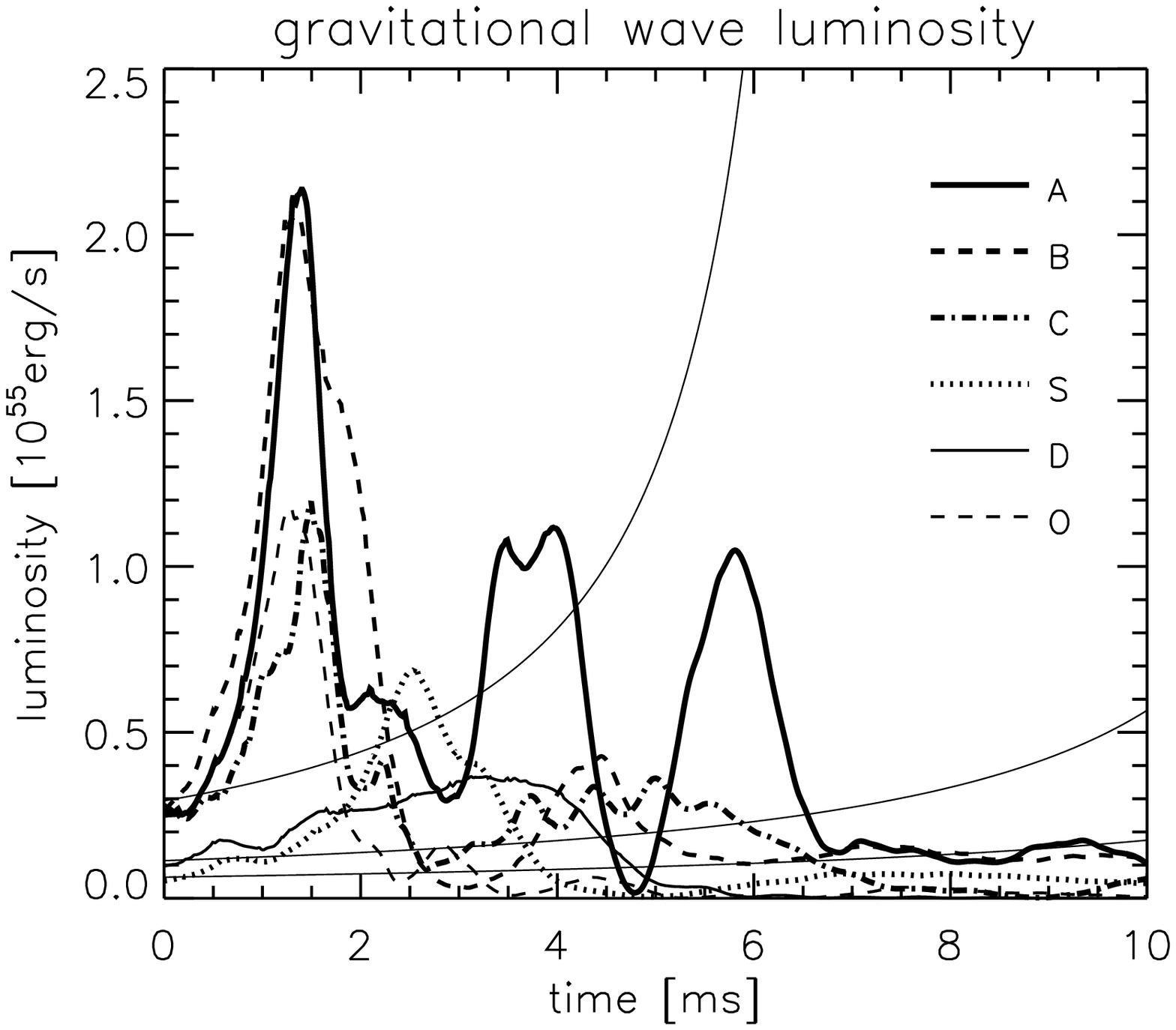} &
%  \put(-0.92,6.75){{\Large \bf \sf c}}  &
  \epsfxsize=8.8cm \epsfclipon \epsffile{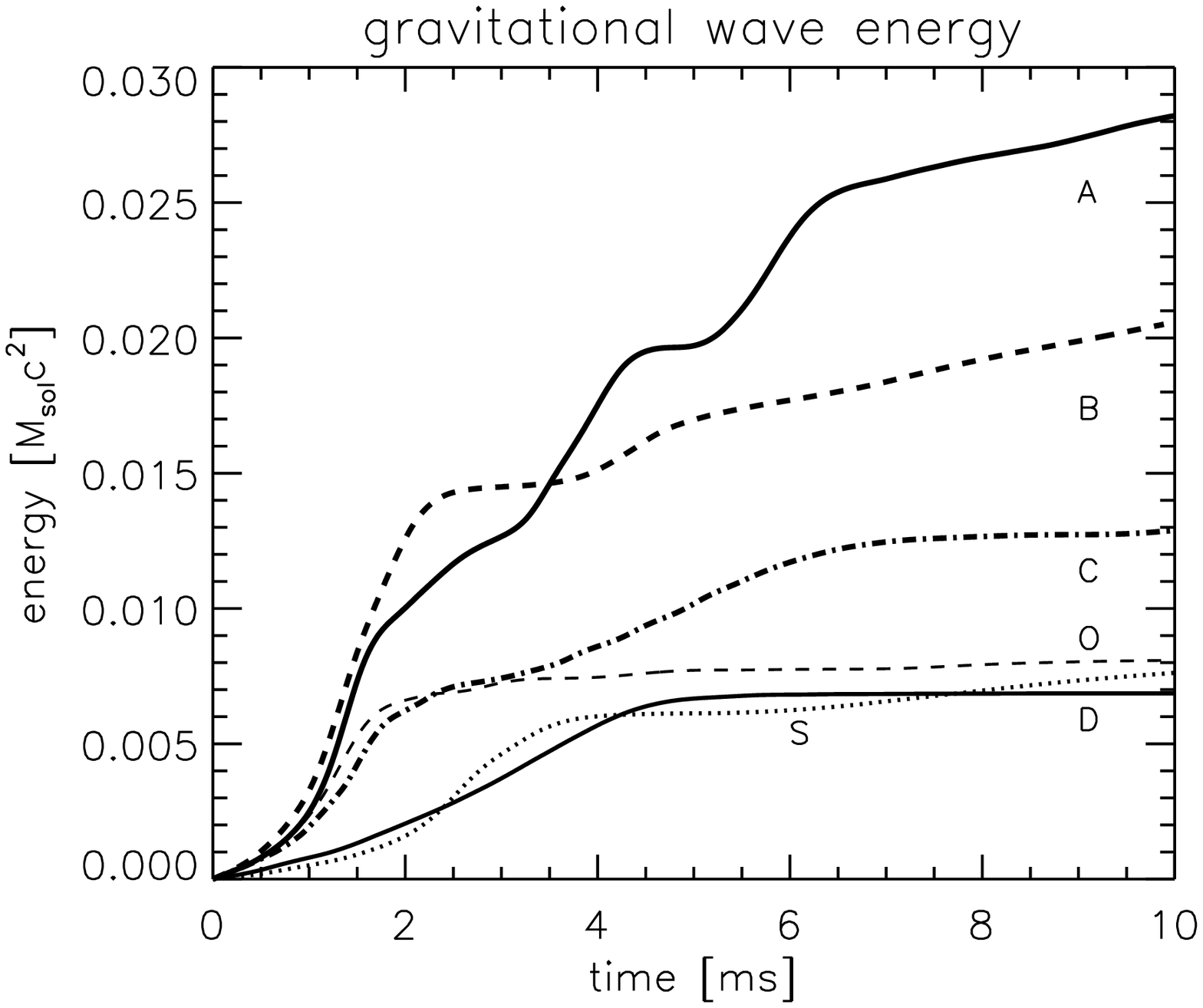} \\[-2ex]
%  \put(-0.92,6.75){{\Large \bf \sf d}}  \\[-4ex]
\end{tabular}
\caption[]{
Gravitational-wave luminosity and cumulative energy emitted in gravitational
waves as functions of time for different models. The thin solid curves 
represent the emission of binary systems with two point masses instead of
the neutron stars
\label{fig:9a}
}
\end{figure*}
%********************************** fig.9a ******************************************

%********************************** fig.9 ******************************************
\begin{figure*}[htp!]
\vspace{25cm}
\caption[]{
Gravitational waveforms of Models~A64, B64, C64, O64, S64, and D64. 
Time is measured in milliseconds from the start of the simulations,
the observer is located perpendicular to the orbital plane. The thin
solid lines correspond to the chirp signal of two point masses
\label{fig:9}
}
\end{figure*}
%********************************** fig.9 ******************************************

Before the final plunge, which is marked by the maxima of the gravitational
waveforms, the gravitational wave emission follows the chirp
signal of two point masses rather accurately. We have not started
our simulations with a configuration in rotational equilibrium and 
correspondingly deformed neutron stars. Therefore the neutron stars
oscillate around their new quasi-equilibrium state before the merging,
and the density maximum on the grid shows periodic modulations. Although
these internal pulsations of the two stars are not damped by numerical 
viscosity (which expresses the non-dissipative quality of the code; 
Fig~\ref{fig:8}),
they do not have any visible consequences for the gravitational waveforms.

The deviation from the chirp signal becomes large when the orbital 
instability sets in and the neutron stars get strongly deformed and finally
begin to touch each other. This moment
contains very important information about the properties of the neutron
stars, in particular about their mass-radius relation, which depends on
the stiffness of the nuclear EoS. Similarly important and characteristic
information about the binary system is carried by the post-merging
signal, which is emitted 
if the massive and compact merger remnant is stabilized by pressure
or centrifugal forces and does not collapse to a black hole on a dynamical
timescale. The collection of gravitational-wave amplitudes displayed in 
Fig.~\ref{fig:9} supports this argument. The waveforms for the binary
systems vary strongly with the spins of the neutron stars (compare 
Model~O with Models~A, B, and C) and with the mass ratio of the two stars
(Model~D). Of course, following the post-merging evolution definitely
requires a general relativistic description of the problem, and quantitatively
meaningful predictions of the gravitational-wave emission cannot be made on 
grounds of our basically Newtonian simulations. Relative differences between
the models in dependence of the binary parameters should, however, also 
survive in a relativistic treatment.

Comparison of the gravitational-wave amplitudes of Models~A64, B64, and C64
with results displayed in Fig.~25 of Ruffert et al.~(1996) reveals
discrepancies for the post-merging signals. Because of the better numerical
resolution of our present simulations (the cell size on the finest grid is
0.64~km instead of 1.28~km for the old models), the numerical loss of 
angular momentum is significantly lower now. The influence of 
the resolution can be directly tested
by Models~A32, B32, and C32, which have a factor of two larger cells on the
innermost grid and thus have the same resolution as the old calculations.
The structure of the wave amplitude for Models~A32, B32, and C32 is
therefore much more similar to our previously published results. 

The gravitational-wave luminosity (Fig.~\ref{fig:9a}) peaks when the two
neutron stars plunge into each other and the quadrupole moment of the 
binary changes most rapidly. For Models~A, B, C, O this happens at roughly
the same time, but the maximum luminosity of Models~A and B is about twice
as large as the one of the counterrotating Model~C and of Model~O where the
neutron star spins are in opposite directions. Model~S with smaller neutron 
stars and Model~D with different neutron stars show significantly lower
peak luminosities. For a short time, the luminosity
rises above the value of two orbiting point masses at the same distance.
Because of the finite size of the neutron stars and the rapid development
of a nearly spherical merger remnant, however, the luminosity decreases
quickly after the maximum and never diverges as in case of the arbitrarily 
close approach 
of point masses. In the cases with strong, coherent fluid motion within
the remnant, secondary and even tertiary maxima can occur. Such effects
are absent in Model~D and very weak in Model~O.

Our models are basically Newtonian, therefore the calculated gravitational 
waveforms are of limited usefulness as templates for measurements.
Nevertheless, the simulations can serve for comparison with future, fully
general relativistic models and will help one understanding the properties 
of their gravitational-wave signals.

%********************************** fig.10 ******************************************
\begin{figure*}[htp!]
\begin{tabular}{cc}
  \epsfxsize=8.8cm \epsfclipon \epsffile{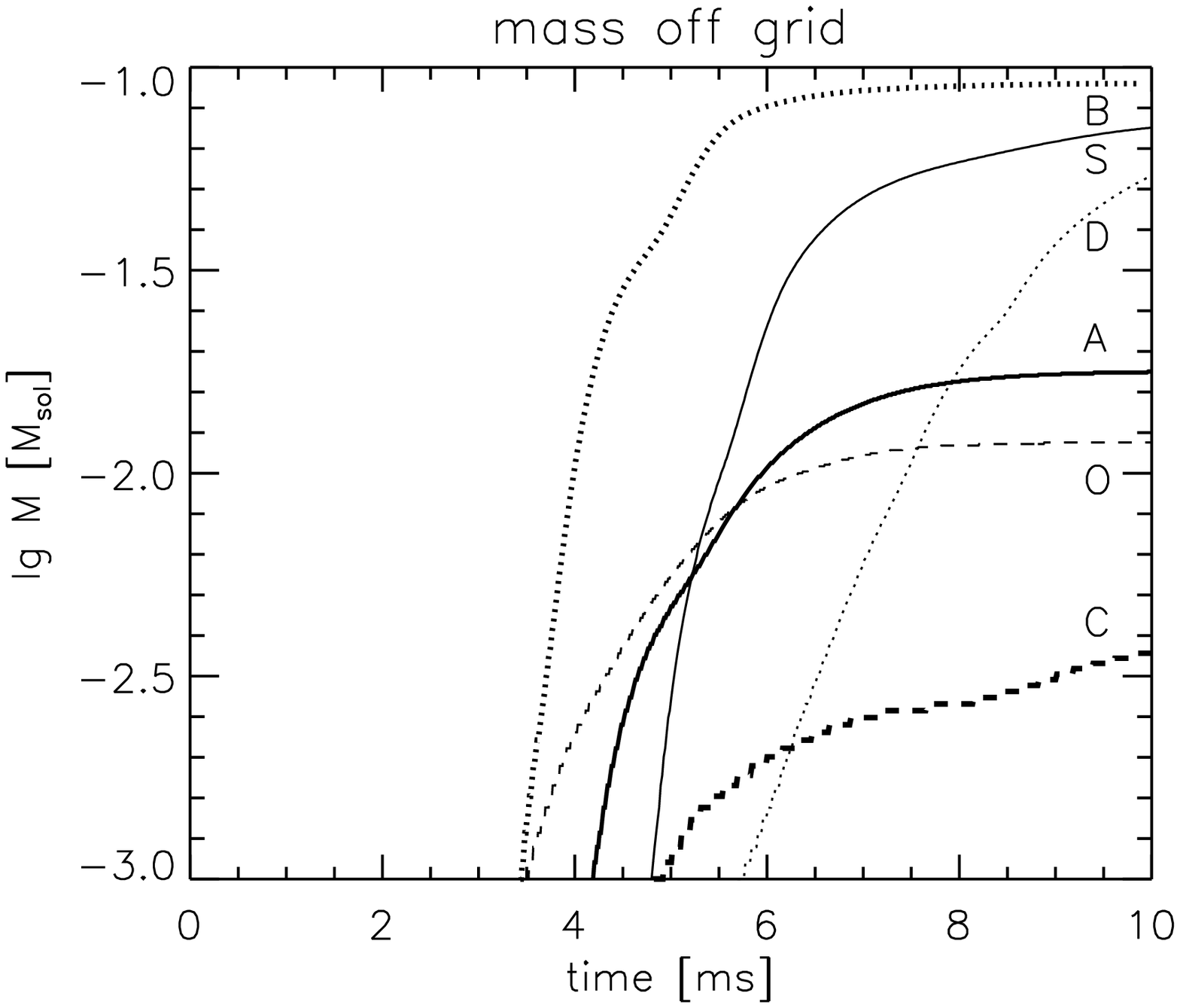} &
%  \put(-0.92,6.75){{\Large \bf \sf c}}  &
  \epsfxsize=8.8cm \epsfclipon \epsffile{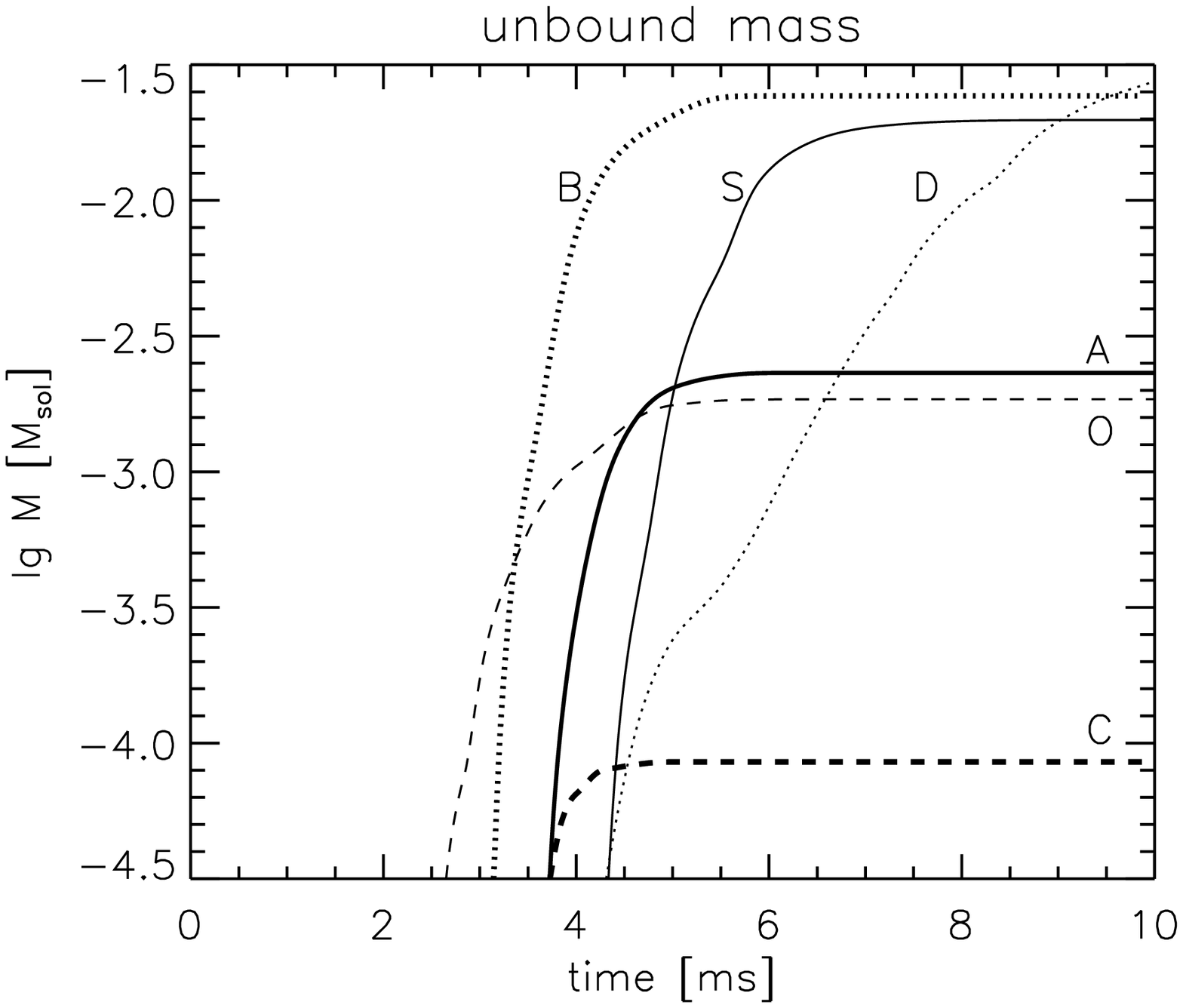} \\[-2ex]
%  \put(-0.92,6.75){{\Large \bf \sf d}}  \\[-4ex]
\end{tabular}
\caption[]{
The cumulative amount of gas that flows off the grid as function of
time for different models (left) and the amount of matter that is
estimated to become unbound, because its total energy (as the sum
of gravitational, kinetic, and internal energies) is positive 
\label{fig:10}
}
\end{figure*}
%********************************** fig.10 ******************************************

\begin{table*}
\label{tab:2}
\caption[]{
Gravitational-wave and neutrino emission properties for all models.
$\widehat{\cal L}$ is the maximum gravitational-wave luminosity,
${\cal E}$ is the total energy emitted in gravitational waves,
$r\widehat{h}$ is the maximum amplitude of the gravitational waves 
scaled with the distance to the observer, $r$,
$L_{\nu_e}$ is the electron neutrino luminosity after approaching
a saturation level at about 8~ms, $L_{\bar{\nu}_e}$ is the
corresponding electron antineutrino luminosity, and
$L_{\nu_x}$ the luminosity of each individual species of heavy-lepton
neutrinos,
$\langle\epsilon_{\nu_e}\rangle$, $\langle\epsilon_{\bar{\nu}_e}\rangle$
and $\langle\epsilon_{\nu_x}\rangle$ are the mean energies of the 
emitted $\nu_e$, $\bar\nu_e$, and $\nu_x$, respectively. $L_\nu$
gives the total neutrino luminosity at the end of the simulation and
$\dot{E}_{\nu\bar{\nu}}$ denotes the integral rate of energy deposition by
neutrino-antineutrino annihilation.
}
\begin{flushleft}

\begin{tabular}{lccccccccccc}
\hline\\[-3mm]
model &   $\widehat{\cal L}$ & ${\cal E}$ & $r\widehat{h}$ &
   $L_{\nu_e}$ & $L_{\bar{\nu}_e}$ & $L_{\nu_x}$ & $L_\nu$ & 
   $\langle\epsilon_{\nu_e}\rangle$ & 
   $\langle\epsilon_{\bar{\nu}_e}\rangle$ & 
   $\langle\epsilon_{\nu_x}\rangle$ & 
   $\dot{E}_{\nu\bar{\nu}}$  \\
 & {\scriptsize$10^{55}\frac{\rm erg}{\rm s}$} &
   {\scriptsize$10^{52}$ erg} & {\scriptsize$10^4$cm} &
   {\scriptsize$10^{53}\frac{\rm erg}{\rm s}$} &
   {\scriptsize$10^{53}\frac{\rm erg}{\rm s}$} &
   {\scriptsize$10^{53}\frac{\rm erg}{\rm s}$} &
   {\scriptsize$10^{53}\frac{\rm erg}{\rm s}$} &
   {\scriptsize MeV} & {\scriptsize MeV} & {\scriptsize MeV} & 
   {\scriptsize$10^{50}\frac{\rm erg}{\rm s}$}
\\[0.3ex] \hline\\[-3mm]
A32 & 2.3 & 4.9 & 8.6 
    & 0.8  & 2.0  & 0.43 & 4.5 & 12. & 18. & 27. & --- \\
A64 & 2.1 & 5.2 & 8.6 
    & 0.9  & 2.6  & 0.37 & 5.0 & 11. & 17. & 27. & 90.5 \\[0.7ex]
B32 & 2.2 & 3.8 & 9.2 
    & 0.6  & 2.3  & 0.45 & 4.7 & 12. & 17. & 25. & --- \\
B32w& 2.2 & 3.5 & 8.4
    & 0.65 & 1.9  & 0.23 & 3.5 & 13. & 15. & 25. & --- \\ %Grbt
B32w'&  2.2 & 3.5 & 8.4
    & 0.39 & 1.3  & $>$0.08 & 2.0 & 11. & 15. & 22. & --- \\ %Grbp
B32c& 2.1 & 3.5 & 8.4
    & $>$0.45 & 1.5 & $>$0.09 & $>$2.3 & 11. & 15. & 22. & --- \\  %Grbf
B32$^5$c & 2.1 & 3.7 & 8.3
    & 0.65 & 1.0 & 0.08 & 1.9 & 12. & 16. & 23. & --- \\  %Grbl
B64 & 2.1 & 3.7 & 8.9 
    & 0.6  & 1.8  & 0.22 & 3.3 & 12. & 17. & 25. & 70.0 \\
B64c& 2.1 & 3.6 & 8.5
    & $>$0.62 & $>$1.6& $>$0.07 & $>$2.5 & 13. & 16. & 24. & --- \\[0.7ex] %GrbF
C32c& 0.85 & 3.4 & 6.0
    & $>$1.0 & $>$2.2 & $>$0.2 & $>$4.0 & 11. & 16. & 27. & --- \\  %Grbg
C64c& 1.2  & 2.3 & 6.0
    & $>$1.1 & 2.3  & 0.13 & $>$4.0 & 11. & 16. & 26. & --- \\[0.7ex]  %GrbG
O32 & 1.3 & 1.5 & 6.9
    & 0.4  & 0.9  & 0.02 & 1.4 & 12. & 19. & 23. & --- \\
O64 & 1.2 & 1.5 & 6.9
    & 0.4  & 1.1  & 0.02 & 1.6 & 14. & 17. & 24. & --- \\[0.7ex]
S32 & 0.65& 1.4 & 5.2 
    & 0.4  & 1.2  & 0.18 & 2.3 & 11. & 16. & 25. & --- \\
S64 & 0.70 & 1.4 & 5.5 
    & 0.3  & 0.9  & 0.07 & 1.5 & 11. & 16. & 25. & --- \\[0.7ex]
D32 & 0.46 & 0.96& 5.8 
    & 0.4  & 1.4  & 0.19 & 2.6 & 12. & 17. & 24. & --- \\
D64 & 0.37 & 1.25& 5.5 
    & 0.4  & 1.0  & 0.16 & 2.0 & 12. & 19. & 26. & --- \\[0.7ex]
\hline
\end{tabular}
\end{flushleft}
\label{tab:models2}
\end{table*}

\subsection{Dynamical mass loss} 

Immediately after the merging of the neutron stars, tidal arms begin
to reach out through the outer Lagrange points of the binary system.
A little later, between 3 and 6 milliseconds after the start of the 
simulations, these spiral arms become inflated, because the   
matter expands with radial velocities up to one third of
the speed of light. At that time the spiral arms reach the boundaries of
the computational volume and mass flows off the grid in our
simulations. This phase coincides roughly with the moments shown in the 
middle panels of Figs.~\ref{fig:2}--\ref{fig:6} and can be recognized in
Fig.~\ref{fig:10} from the steep increase of the plotted curves.
We monitor the fraction of the matter which fulfills the condition that its
total specific energy as the sum of gravitational, kinetic and internal 
energies is positive. This matter is considered to potentially become unbound.

The amount of matter which is stripped off the tips of the spiral arms depends
on the total angular momentum of the coalescing binary system.
For solid-body type initial rotation we find the largest values: Roughly
a tenth of a solar mass leaves the grid, and about 20--30\% of this matter 
might escape the system. The mass ejection is roughly a factor of ten smaller
for the cases of irrotational neutron stars and neutron star spins in opposite
directions, and another factor of ten smaller for the configuration with
initially counterrotating stars (Table~\ref{tab:models1} and Fig.~\ref{fig:10}).

The gas mass $M_{\mathrm{g}}$ that is swept off the grid and the mass 
$M_{\mathrm{u}}$ that may get unbound 
exhibit only a rather weak dependence on the grid resolution, which can be
seen by comparing models with different choices of grids in 
Table~\ref{tab:models1}. Of course,
the size of the largest grid has a significant influence on $M_{\mathrm{g}}$,
but again does not affect $M_{\mathrm{u}}$ very much (Model~B32$^5$c).

Although the correlation of mass ejection with the angular momentum of
the binary system appears plausible, and basically is consistent with the
results of Rosswog et al.~(2000), we cannot claim to have determined
final values for the dynamical mass loss. There are several factors of 
uncertainty which affect our numbers. The finite environmental gas density,
which we have to assume around the neutron stars for numerical reasons, has
probably negligible dynamical influence. More problematic is the decreasing
resolution of the nested grids towards the boundaries of our computational 
volume. The largest uncertainty, however, is connected with the fact that 
we cannot follow the expanding matter until it moves ballistically. Therefore
it is unclear how efficiently internal energy is finally converted into bulk 
kinetic energy by hydrodynamical effects (i.e., $P{\mathrm{d}}V$-work).
Our results for the ejected mass can only be considered as best estimates,
supported by the fact that the expansion of the tidal tails is mainly a 
kinematic effect and only to a minor degree driven by pressure forces.

%********************************** fig.11 ******************************************
\begin{figure*}[htp!]
\vspace{16cm}
\caption[]{
Electron fraction $Y_e$ in the orbital plane at two different times
for Model~B64 (upper row) and Model~D64. The contours are spaced linearly
in steps of 0.02, 
beginning with a minimum value of 0.02, which is also adopted for
the very dilute environmental medium.
The bold lines correspond to values of 0.02, 0.06, 0.1, 0.16, 0.2, and 0.3.
The maximum values are 0.28 in the upper left plot, 0.18 in the upper right
one, 0.35 in the lower left one, and 0.33 in the lower right one
\label{fig:11}
}
\end{figure*}
%********************************** fig.11 ******************************************

Matter which is ejected during the merging of binary neutron stars may have 
important implications for the production of heavy elements in our Galaxy. Dynamical
events involving neutron stars (mergers, explosions) have been speculated to be 
possible sources of r-process nuclei, both because of the high neutron densities
present in the interior of neutron stars and because of the neutron-rich nuclei 
which exist in the crust below the surface of the cold neutron star 
(e.g., Hilf et al.~1974; Lattimer \& Schramm 1974, 1976; Lattimer et al.~1977;
Symbalisty \& Schramm 1982; Eichler et al.~1989; 
Meyer 1989; Freiburghaus et al.~1999). The crust below a density of about
$2\times 10^{14}\,$g$\,$cm$^{-3}$ (see, e.g., Weber 1999) can contain up to
several hundredths of a solar mass for a neutron star with a gravitational mass
around $1.4\,M_{\odot}$ (e.g., Akmal et al.~1998), although the exact value
is somewhat EoS dependent. With an estimated
merger rate of one per $10^{5\pm 1}$~years and a mean mass loss per event
of $\sim 10^{-3}$ solar masses, about 10--1000 solar masses of r-process
material might have been produced by $10^4$ to $10^6$ such events over the 
history of our Galaxy.

Quantitative predictions of the abundance yields are not easy to obtain.  
Ideally, the nuclear reactions and beta decays of neutron-rich isotopes 
should be followed along with the hydrodynamic flow of the expanding
ejecta, because heating by beta decays and neutrino losses can influence 
their thermal and chemical evolution. Moreover, it is essential to start the
nucleosynthesis calculations with appropriate initial conditions. This concerns
the initial temperature, initial electron fraction $Y_e = n_e/n_{\mathrm{b}}$,
and the initial nuclear composition of the surface material of cold neutron
stars. 

A first attempt for a nucleosynthetic evaluation of the merger ejecta was
performed as a post-processing step of hydrodynamical data (Freiburghaus et al.~1999).
Instead of being
self-consistently determined as the consequence of neutrino reactions, the electron
fraction was set to a chosen value. In addition, the initial temperature of the 
ejected neutron star matter was considered to be high enough ($T\ga 5\times 10^9$~K)
for the composition to be determined by nuclear statistical equilibrium (NSE). This is
not necessarily a correct assumption, because the matter that is flung out in the
tidal arms is not subject to any heating processes which could significantly 
increase the surface temperature of the originally cold neutron stars. In 
three-dimensional hydrodynamical simulations of neutron star mergers, however,
the limited numerical resolution does not allow one to accurately trace the thermal 
history of such cold matter, but numerical viscosity und numerical noise lead to
artificial heating (for our code, such problems
are discussed in Sects.~\ref{sec:numerics} and \ref{sec:simulations}). If the
temperature stays low, which is most likely the case in the ejected and decompressed
matter, the initial nuclear composition of the neutron star crust is not erased by 
the onset of NSE. Instead, the neutron-rich nuclei start to undergo beta decays
and the final distribution of stable or long-living nuclei is sensitive to the 
initial composition (e.g., Hilf et al.~1974, Lattimer et al.~1977,
Meyer 1989, Sumiyoshi et al.~1998).

In our simulations the electron fraction evolves as a result of the emission
of electron neutrinos $\nu_e$ and antineutrinos $\bar\nu_e$. Near the tips of
the tidal tails, however, the neutrino producing reactions are not fast enough to
cause a sizable change of the initial value of $Y_e$ on the short timescales of
the dynamic expansion. This is so because the
temperature near the neutron star surface was chosen to be ``only'' around
1--2~MeV in the beginning (see Fig.~2 in Ruffert et al.~1996; such values of the
temperature are certainly unphysically high for cold initial stars, but they can 
be considered as low for the present discussion) and the temperature does not
increase in the spiral arms during the subsequent evolution
(Figs.~\ref{fig:2}--\ref{fig:6}). As a consequence, $Y_e$ can serve as
a tracer quantity, 
and our results indicate that only matter that was initially located
close to the surfaces of the original stars is ejected during coalescence.
This matter essentially
retains its initial, very neutron-rich state with $Y_e$ 
significantly less than 0.1 (which corresponds
to the neutrinoless beta-equilibrium (or beta-stable) state
of cold matter in the crust below the very 
thin envelope of a neutron star). The typical
electron fraction of the ejecta is found to be around 0.02--0.04
(Fig.~\ref{fig:11}), much lower than
determined by Freiburghaus et al.~(1999) as favorable for an r-processing that  
leads to a solar-like abundance distribution. 
We remind the reader here, however, that the original EoS by
Lattimer \& Swesty (1991) does not provide values below $Y_e = 0.03$,
and we use a simple extrapolation of the EoS in this regime. 
In fact, the exact value of $Y_e$ in the neutron star crust is not well
known because of uncertainties in the physics, for example
associated with the nucleon symmetry energy or with phase transitions 
due to isospin states in case of very low $Y_e$.
In addition,
the current grid zoning of our three-dimensional models does not allow us
to resolve the thin crust and envelope of the cold initial neutron stars, where 
the state of neutrinoless beta-equilibrium corresponds to a positive gradient
of $Y_e$ and thus higher values of $Y_e$ towards the surface.

%********************************** fig.11b *****************************************
\begin{figure}
  \epsfxsize=9.2cm \epsfclipon \epsffile{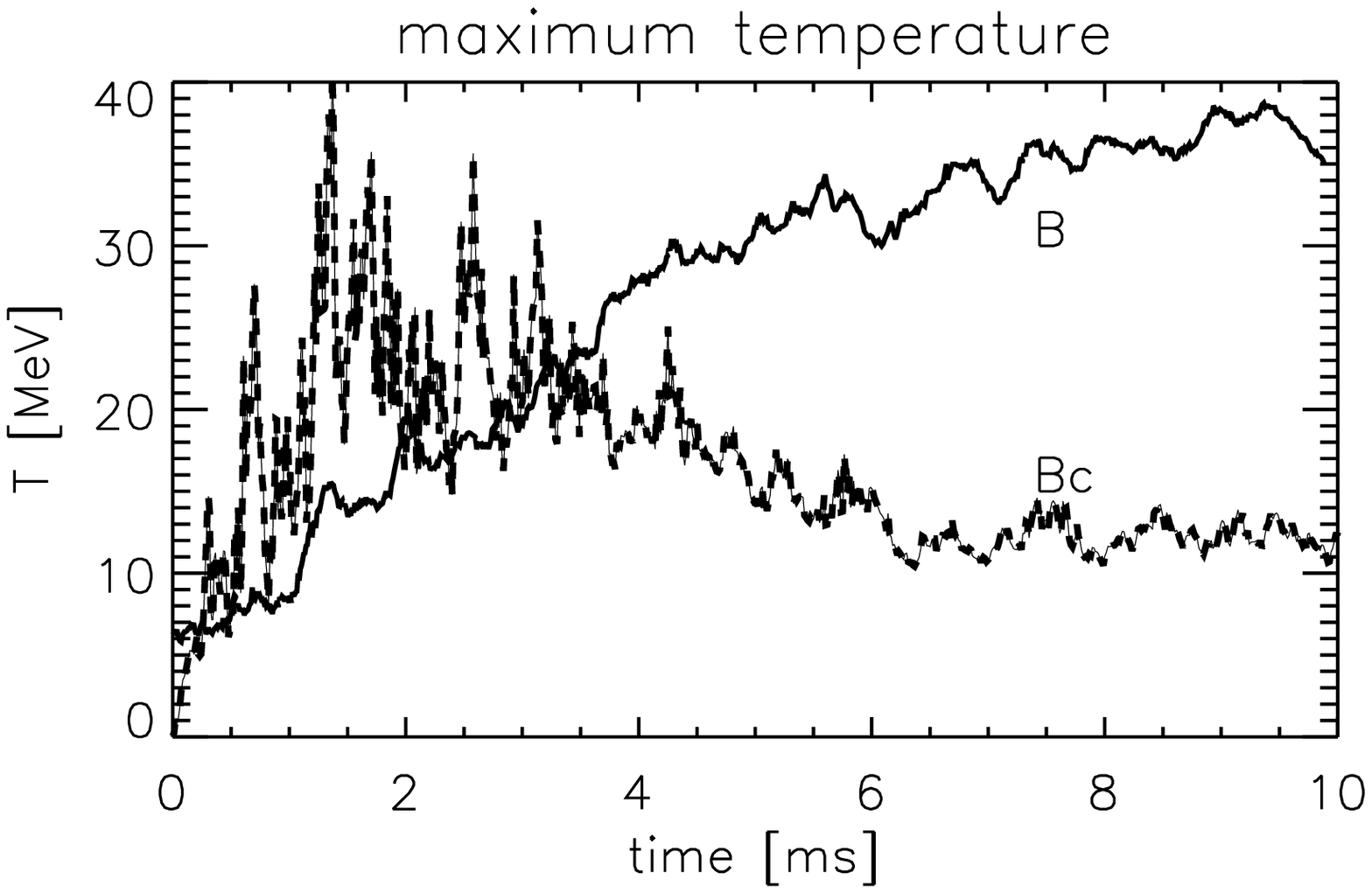}
  \caption[]{Maximum temperature on the grid as a function of time for
   Models~B64 and B64c}
\label{fig:11b}
\end{figure}
%********************************** fig.11b *****************************************

The question of dynamical mass loss is not unequivocally answered on grounds of our
simulations or those of Rosswog et al.~(1999, 2000). A source of uncertainty is
the nuclear EoS which determines the structure and properties of 
the merging neutron stars and the dynamics of the merging process.
Rosswog et al.~(2000) found that the mass loss is sensitive to the stiffness of
the EoS. In particular the conditions in the supranuclear regime are highly
uncertain. EoSs which have been developed for supernova simulations and are
particularly suitable for hydrodynamic modeling as described here
(because they provide all the required information in essentially all interesting
regions of the parameter space),
in particular the EoS of Lattimer \& Swesty (1991) and Shen et al.~(1998),
do actually not include a detailed microphysical description of the regime 
beyond about twice the density of the nuclear
phase transition. New hadronic degrees of freedom (e.g., a phase with pions, 
kaons or hyperons) or quark matter could be present there and would soften the EoS.
While probably not crucial for stellar core-collapse and supernova 
simulations, this density regime determines the cooling evolution of new-born
neutron stars and should also affect the merging of neutron stars. For example,
if the supranuclear EoS is sufficiently soft, the maximum mass of stable 
(nonrotating) neutron stars can be as low as 1.5--1.6$\,$M$_{\odot}$. This possibility
cannot be excluded, neither on grounds of theoretical calculations of the state of
matter at supranuclear densities (e.g., Weber 1999, Heiselberg \& Pandharipande 2000), 
nor on
grounds of observed neutron star masses. Although rapid and differential rotation  
can have a significant stabilizing influence (Baumgarte et al.~2000), 
such a soft EoS would probably not allow the merger remnant to escape the 
collapse to a black hole on a dynamical timescale (Shibata \& Ury\=u 2001). 
Therefore the question will have to be investigated in more detail
by future general relativistic simulations, whether in this case
mass ejection, which occurs with some time delay after the final plunge, can still
take place.

%********************************** fig.12a ******************************************
\begin{figure*}[htp!]
\vspace{16cm}
\caption[]{
Temperature and electron fraction in the orbital plane of Model~B64c 
for two different times. In this model the entropy equation was used 
to follow the temperature evolution. The temperature 
contours are spaced linearly with levels at 1, 2, 4, 6, 8 and 10~MeV,
the $Y_e$ contours are given in steps of 0.02, starting with a value
of 0.02. The plots should be compared with the corresponding panels 
in Figs.~\ref{fig:3} and \ref{fig:11}
\label{fig:12a}
}
\end{figure*}
%********************************** fig.12a ******************************************

%********************************** fig.12 ******************************************
\begin{figure*}[htp!]
\vspace{14.cm}
\caption[]{Mean energies (left) and luminosities (right)
of $\nu_e$ (label ``e''), $\bar\nu_e$ (label ``a'') and heavy-lepton neutrinos
and antineutrinos, $\nu_x$ (label ``x''), as functions of time for 
Models~B64 (top) and B64c (bottom). The luminosity of $\nu_x$ includes 
contributions from muon and tau neutrinos and antineutrinos
\label{fig:12}
}
\end{figure*}
%********************************** fig.12 ******************************************

%********************************** fig.12b ******************************************
\begin{figure*}[htp!]
\vspace{16cm}
\caption[]{
Total neutrino energy loss rates (in erg$\,$cm$^{-3}$s$^{-1}$) from the
matter in the orbital plane and perpendicular to the orbital plane
for Model~B64 (top) and Model~B64c (bottom) at the end
of the computed evolution. The plots show the central region of the
computational grid with the neutrinospheres of $\nu_x$, $\bar\nu_e$, and
$\nu_e$ marked by dashed lines (in this order from inside outward). The
contours are spaced logarithmically in steps of 0.5~dex
\label{fig:12b}
}
\end{figure*}
%********************************** fig.12b ******************************************

%********************************** fig.12c ******************************************
\begin{figure*}[htp!]
\begin{tabular}{cc}
  \epsfxsize=8.8cm \epsfclipon \epsffile{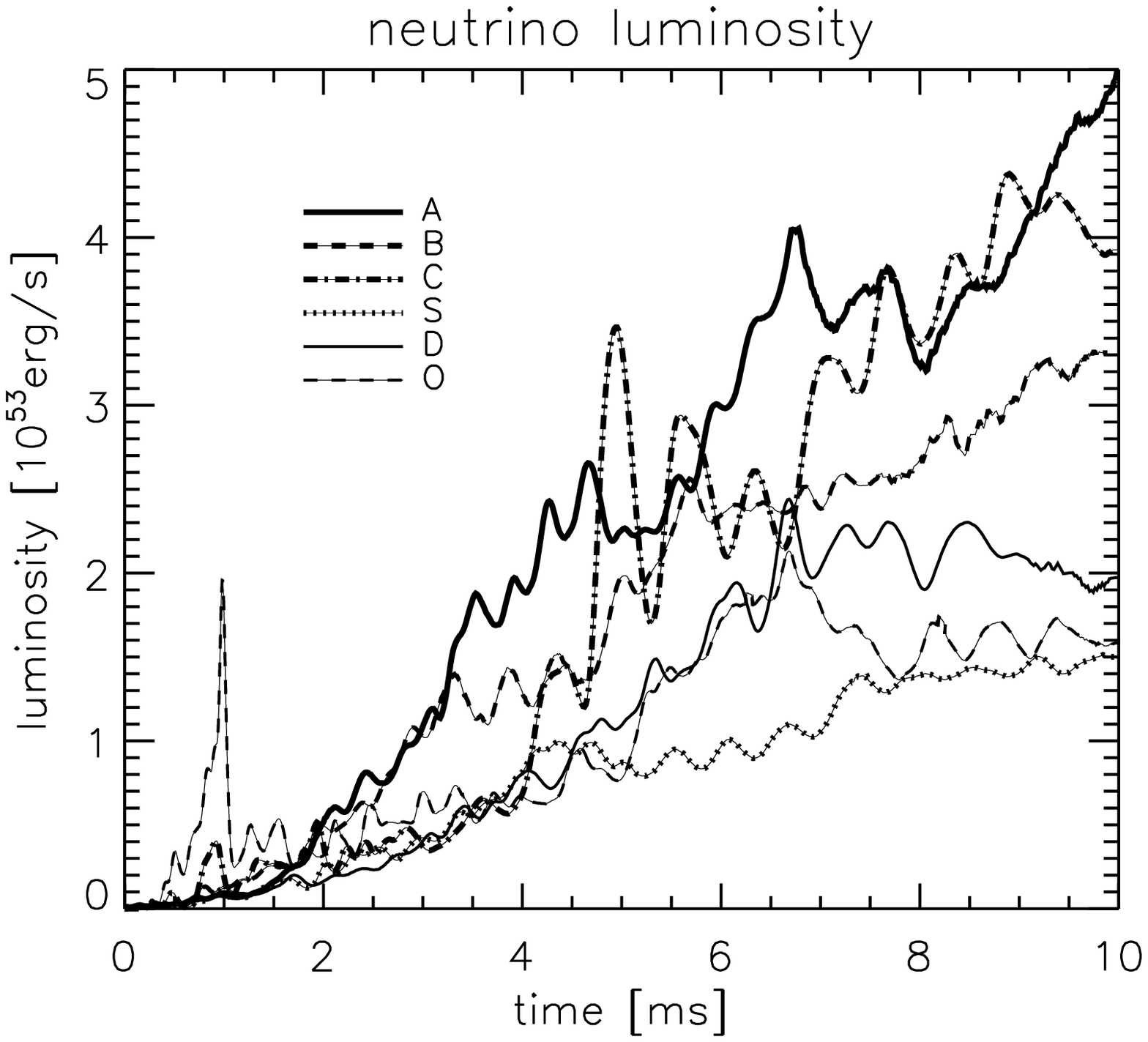} &
%  \put(-0.92,6.75){{\Large \bf \sf c}}  &
  \epsfxsize=8.8cm \epsfclipon \epsffile{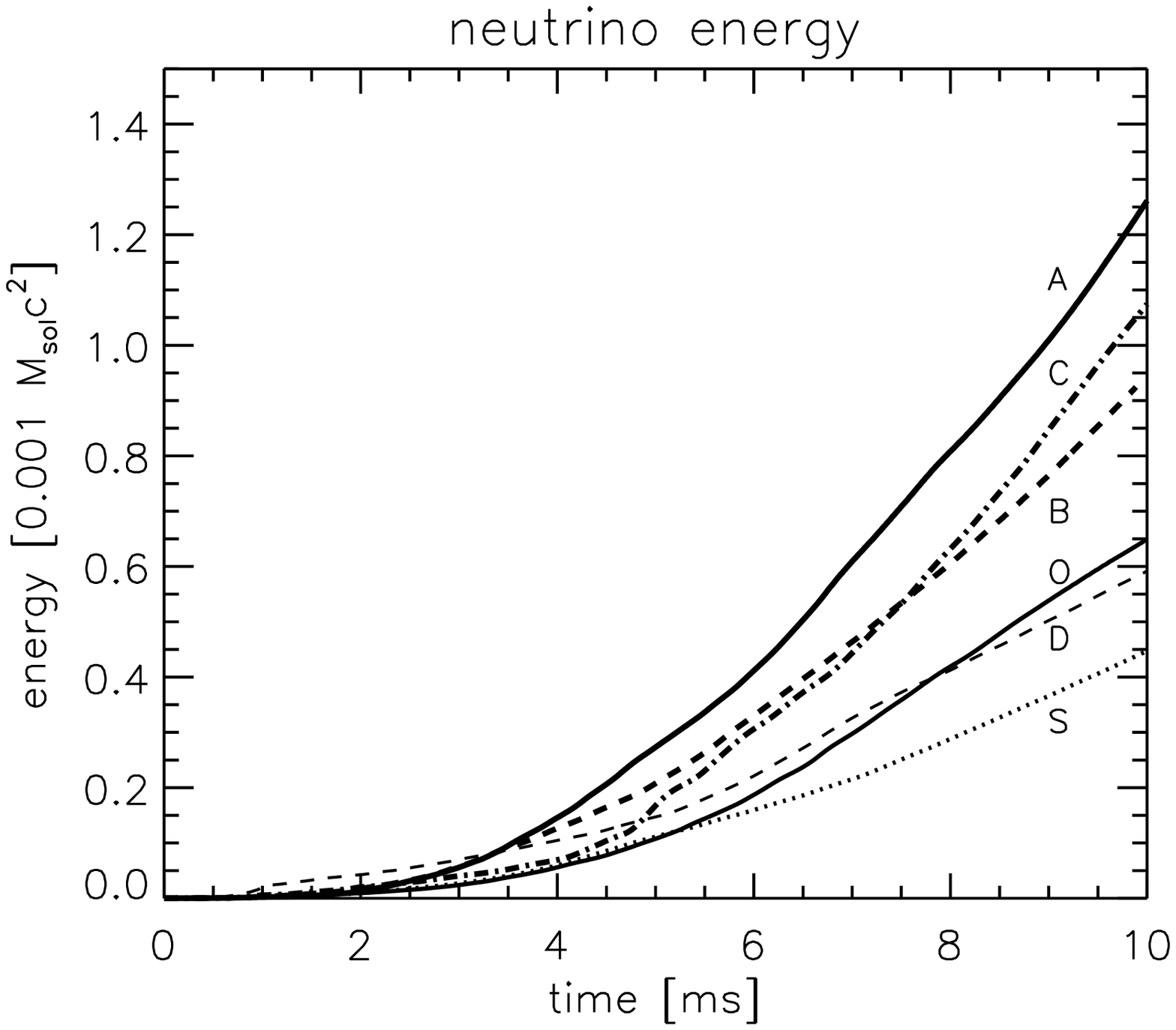} \\[-2ex]
%  \put(-0.92,6.75){{\Large \bf \sf d}}  \\[-4ex]
\end{tabular}
\caption[]{
Total neutrino luminosity (summed up for all flavors) and cumulative energy
emitted in neutrinos as functions of time for different models
\label{fig:12c}
}
\end{figure*}
%********************************** fig.12c ******************************************

%********************************** fig.12d ******************************************
\begin{figure*}[htp!]
\vspace{16cm}
\caption[]{
Neutrino energy loss rates (in erg$\,$cm$^{-3}$s$^{-1}$) from the
matter in the orbital plane of Model~B64 at the end of the computed
evolution. The rates for $\nu_e$ (top left), $\bar\nu_e$ (top right),
the sum of muon and tau neutrinos and antineutrinos (bottom left) and
for neutrinos and antineutrinos of all flavors are given logarithmically,
with steps of 0.5~dex. The dashed lines in the last plot indicate the
positions of the neutrinospheres in the orbital plane
\label{fig:12d}
}
\end{figure*}
%********************************** fig.12d ******************************************

\subsection{Thermal evolution and neutrino emission}

Computing the temperature evolution is crucial for calculating the 
neutrino emission of the merging neutron stars. Unfortunately, there are
a number of numerical problems which affect the accuracy of the temperature
determination. For our code and simulations, these problems have already 
been addressed in Sects.~\ref{sec:numerics} and \ref{sec:simulations}.

When doing simulations with a finite difference
scheme, the need to assume a medium with a finite density on the
computational grid constitutes a problem.
In a non-rotating frame of reference (or a frame of
reference which does not move with the same angular velocity as the stars),
the neutron stars move through this dilute medium with a high relative 
velocity. This leads to shock heating at the forward 
surfaces of the neutron stars. 
We reset the temperature to lower values after each time step 
for densities below a certain threshold value so that only
dilute material is affected by this procedure.
We also detect artificially heated, isolated
grid zones and reduce the local temperature by averaging over neighbouring,
well behaved zones. Since only relatively few cells with a rather low density
are involved, this procedure does not introduce a noticable violation of the
conservation of the total energy during the simulations (Fig.~\ref{fig:8b}). 
Moreover, the affected volume and mass decrease when the resolution is improved.

Another, more severe problem is connected with the fact that cool neutron 
star matter is very degenerate, i.e., the thermal energy contributes only
a minor fraction to the internal energy. Since our hydrodynamics code
computes the specific internal energy, from which the temperature is 
determined, as the difference between internal plus kinetic energy of the fluid 
and its kinetic energy, any small error introduced there leads to sizable 
perturbations in the temperature. The thus calculated temperature cannot be
very accurate. For this reason, we decided to repeat some of our simulations
by using an additional entropy equation, which allows us to follow the 
thermal history without the described sources of noise 
(see Sect.~\ref{sec:numerics} for more details). Note that the entropy
equation serves only for computing the temperature, but does not replace the
energy equation as one of the conservation laws which are solved for describing
the hydrodynamical flow. The neutrino source terms in the latter equation,
however, are computed with the temperatures as obtained from the entropy
equation.

Models B32 and B32w in Tables~\ref{tab:models1} and \ref{tab:models2}
are two cases which were computed with the same grid resolution, but in the
second model the additional entropy equation was used. Both simulations
were started with nearly the same temperatures in the neutron stars. The
additional Model~B32c represents a case where the initial central temperature 
was chosen to be as low as 0.05~MeV instead of
6.76~MeV in Model~B32 or 5.44~MeV in Model~B32w. The values for the 
different entries in the tables show that the dynamic quantities and
the gravitational-wave emission are nearly the same. The quantity that 
is most sensitive to the different treatments is the mass $M_{\mathrm{u}}$
which can become unbound from the system. Despite of the different 
thermal evolutions, even the overall properties of the neutrino emission
at the end of the computations are rather similar.

For discussing more details,
let us now consider two better resolved calculations, Models~B64
and B64c. These two exemplary models differ by the
use of the entropy equation in combination with a low initial
central temperature in Model~B64c, namely 0.05~MeV instead of
6.25~MeV in the neutron stars of Model~B64 (Table~\ref{tab:models1}).
The lower initial temperature is more realistic than our usually
``warm'' initial conditions, because viscous heating prior to the
merging is unlikely to achieve temperatures in excess of $10^9\,$K
(Bildsten \& Cutler 1992, Kochanek 1992, Lai 1994).
Starting with a cold initial state requires the use of the entropy
equation (see above).

Figure~\ref{fig:11b} shows the maximum temperatures 
on the grid as functions of time for Models~B64 and B64c.
Indeed, the maximum temperatures 
evolve significantly differently. This is confirmed by the upper panels
of Fig.~\ref{fig:12a}, which show the temperature in the equatorial plane
of Model~B64c for two times, which can directly be compared with snapshots
given for Model~B64 in Fig.~\ref{fig:3}. Note that in Model~B64 the maximum
temperature during the early phase is limited by the spike-correction 
procedure, whereas such a temperature reduction is not applied to the 
``entropy-temperature''' in Model~B64c.

The different temperatures in both simulations affect the neutrino 
source terms for lepton number and energy, which are very sensitive to
the gas temperature (see the Appendices in Ruffert et al.~1996). Since the
neutrino source terms enter the hydrodynamics equations and the continuity 
equation for the electron lepton number, they could in principle have
consequences for the dynamical evolution of the models. In reality, however,
the neutrino emission is irrelevant for the dynamics, because the associated
energy is too small on the short timescale of the calculation.

Although Model~B64c has higher peak temperatures during the early phase
of the evolution, the average temperatures are considerably lower than in
Model~B64 (Figs.~\ref{fig:3} and \ref{fig:12a}).
These lower temperatures reduce the (anyway small)
changes of the lepton number due to the neutrino emission in the expanding 
tidal tails. 
The panels in the second row of Fig.~\ref{fig:12a} confirm that the 
lepton number remains essentially unchanged from its value of about 0.02
near the surfaces of the original neutron stars (compare also with 
Fig.~\ref{fig:11}).

The temperature distributions in the orbital plane of Models~B64 and
B64c at the end of the simulations (see Figs.~\ref{fig:3} and \ref{fig:12a})
reveal that only moderate differences occur in the cloud of low-density
material ($\rho \la 10^{12}$--$10^{13}\,$g$\,$cm$^{-3}$ at $r \ga 50$~km)
which surrounds the much denser and very compact core of the merger
remnant. Shock heating has increased the temperatures in this cloud to
values between 1~MeV and 5--6~MeV in both models. Much more pronounced 
differences between the models can be found in the interior of the 
compact core, where Model~B64 is significantly hotter ($T > 10$--15~MeV), 
mainly due to the dissipation of kinetic energy by numerical viscosity. The 
corresponding heating depends on the shear motions, which are particularly
strong during the phase when the neutron stars merge, and later during the
numerous revolutions of the rapidly spinning central core within the 
surrounding layer of gas.

Most of the neutrino emission comes from the outer regions of dilute
gas, because the dense core of the remnant is much less
transparent to neutrinos. For this reason, the luminosities and
mean energies of the emitted neutrinos and antineutrinos of all flavors 
are rather similar for Models~B64 and B64c (Fig.~\ref{fig:12}), in 
particular towards the end of the simulations, when the extended gas 
cloud has reached a quasi-steady state. Before the tidal arms
expand and shock heating and viscous shear has raised the temperatures
in the outer parts of the merged stars, the neutrino emission 
of the initially cold Model~B64c in fact stays on a relatively low level 
($\sim 10^{52}\,$erg$\,$s$^{-1}$). During this phase the mean
energies of the emitted neutrinos show large fluctuations.

The similarity of the neutrino production in the late phase of Models~B64 
and B64c is also visible in Fig.~\ref{fig:12b}, where the total loss rates
of neutrino energy per unit volume are displayed for both models. One 
can see that the dominant contribution to the emission of neutrinos of all 
flavors stems from the outer parts of the merger remnant and not from the
very dense core. In both models, the location
of the neutrinospheres, defined where the optical depth 
perpendicular to the orbital plane (i.e., effectively the minimum in 
all three coordinate directions of the cartesian grid) drops below unity
(Eq.~7 in Ruffert \& Janka~1999), is also very
similar. In the orbital plane the neutrinosphere radius is around 60--70~km.
Perpendicular to the equator the density gradient is very steep and
the neutrinospheres reach up to a vertical distance of about 20~km. 
The neutrinospheres of electron neutrinos ($\nu_e$), electron antineutrinos
($\bar\nu_e$) and of the heavy-lepton neutrinos 
($\mu$ and $\tau$ neutrinos and antineutrinos, $\nu_x$,
are produced with the same rates and ``see'' essentially the same opacity) 
are very close to each other. The neutrinosphere of $\nu_x$ is
slightly smaller because there is no contribution of charged-current
neutrino-nucleon interactions to the $\nu_x$ opacity.
 
The typical total neutrino luminosities at the end of the computed 
evolution are of the order of a few times $10^{53}\,$erg$\,$s$^{-1}$
(Fig.~\ref{fig:12c}). Within about 10$\,$ms the
models radiate an energy equivalent of roughly $10^{-3}\,M_{\odot}$ in
neutrinos, which is more than an order of magnitude less than in
gravitational waves (compare Figs.~\ref{fig:12c} and \ref{fig:9a}). 
These neutrino luminosities and energies are considerably larger than 
those given by 
Ruffert et al.~(1997a) for comparable models. The reason for this result
is the use of the much larger grid in the current simulations. The 
previous models 
suffered from the problem that a significant amount of matter left 
the computational volume when the tidal tails expanded. In the present
calculations this gas is
wrapped up to form the shock-heated envelope around the dense core. 
Note that the position of the neutrinospheres is therefore
{\em outside} of the computational volume that was employed by 
Ruffert et al.~(1997a).

The largest contribution to the neutrino emission comes from electron 
antineutrinos, because positron captures on free neutrons dominate
electron captures on protons. In the decompressed and hot neutron
star matter, which forms the envelope around the merger core, 
neutrons are very abundant and the electron degeneracy is moderate,
for which reason electron-positron pairs are present in large numbers.
Because $\mu$ and $\tau$ neutrinos and antineutrinos are produced
only by pair reactions (mainly $e^- + e^+ \to \nu + \bar\nu$) ---
but not by electron/position captures on protons/neutrons, 
which are the dominant processes for generating $\nu_e$ and $\bar\nu_e$,
respectively --- their
emission is extremely sensitive to the temperature (the energy production
rate scales with $T^9$). Only in very hot models do
they contribute to the neutrino emission on a significant level. For
example, the difference of the total neutrino luminosities of 
Models~B64 and B64c is almost entirely due to the emission
of heavy-lepton neutrinos; the final luminosities of electron neutrinos 
and antineutrinos are essentially the same in both models (Fig.~\ref{fig:12c}).  
The typical mean energies are around 11--13~MeV for the radiated
$\nu_e$, 15--19~MeV for $\bar\nu_e$, and 22--27~MeV for $\nu_x$
(Table~\ref{tab:models2}).

%********************************** fig.13 ******************************************
\begin{figure*}[htp!]
\vspace{8cm}
\caption[]{
Energy deposition rate (in erg$\,$s$^{-1}$cm$^{-3}$) by the annihilation 
of neutrinos and antineutrinos to electron-positron pairs in the surroundings
of the merger remnant for Models~A64 (left) and B64 at the end of the 
computed evolution. Azimuthally
averaged rates are shown in a plane perpendicular to the orbital plane. The
corresponding solid contours are spaced 
logarithmically in steps of 0.5~dex. The dotted lines indicate the 
(azimuthally averaged) isodensity levels, also spaced logarithmically with
steps of 0.5~dex. The rates are evaluated only in regions where the density
is less than $10^{11}\,$g$\,$cm$^{-3}$ 
\label{fig:13}
}
\end{figure*}
%********************************** fig.13 ******************************************

Figure~\ref{fig:12d} gives the energy loss rates per unit volume in
the orbital plane of Model~B64 at the end of the simulated evolution
for all neutrino types ($\nu_e$, $\bar\nu_e$, and $\nu_x$) individually
and for the sum of all contributions. The orbital plane is shown out to
the boundaries of the computational grid. Although the three neutrinospheres
nearly coincide, one notices clear differences of the spatial 
distribution of regions where 
the different types of neutrinos are produced. While $\nu_e$ are rather 
uniformly emitted from most of the matter, there are definite hot spots 
which radiate large amounts of $\bar\nu_e$ and $\nu_x$. This emphasizes
the temperature sensitivity of the corresponding neutrino production
processes, which require the presence of positrons in the stellar medium.

We evaluated two of our models, A64 and B64, for the energy deposition by
the annihilation of neutrinos and antineutrinos to electron-positron pairs,
$\nu + \bar\nu  \to e^- + e^+$,
as described by Ruffert et al.~(1997a) and Ruffert \& Janka (1999). The 
results at the end of the computed evolution (where the
neutrino luminosities are highest) are plotted in Fig.~\ref{fig:13}. 
The largest energy deposition rates per unit volume are found above the
poles of the merger remnant with peak values in excess of 
$10^{32}\,$erg$\,$cm$^{-3}$s$^{-1}$. The numbers for the total rate of
energy deposition in the volume where the gas density is below 
$10^{11}\,$g$\,$cm$^{-3}$ are about $9\times 10^{51}\,$erg$\,$s$^{-1}$ in
case of Model~A64 and $7\times 10^{51}\,$erg$\,$s$^{-1}$ for Model~B64
(Table~\ref{tab:models2}). These rates are more than 20--30 times
bigger than calculated by Ruffert et al.~(1997a). This is mainly explained 
by the much higher neutrino luminosities in our current simulations, which
enter the annihilation rate quadratically. (To some degree the effect 
is also caused by the different geometry, compare Fig.~\ref{fig:13} with 
Fig.~16 in Ruffert et al.~1997a.)

The largest and by far dominant part of the 
energy deposition occurs in the polar regions. Here, however,
also the energy loss rates by neutrino emission reach maxima
(see the panels in the right column of Fig.~\ref{fig:12b}), because
high temperatures are present in a region where the gas is still dense,
but the density gradient is very steep (Fig.~\ref{fig:7}). Therefore 
neutrinos are generated in large numbers and can easily escape to the
transparent regime, as suggested by the fact that the neutrinospheres 
cross the regions of peak emission above the poles of the compact core 
(Fig.~\ref{fig:12b}). With values up to more than
$10^{33}\,$erg$\,$cm$^{-3}$s$^{-1}$, the rate of energy loss exceeds the 
energy input rate by $\nu\bar\nu$ annihilation by more than a factor of 
about 10. Therefore the energy which is transferred to the stellar plasma
will efficiently and immediately be reradiated by neutrino
production processes, and the numbers for the total energy deposition rate
in Table~\ref{tab:models2} overestimate the net heating effect by a large 
factor. 
The conclusion drawn by Ruffert et al.~(1997a) and Janka \& Ruffert (1996) 
remains valid: The energy deposited by neutrino-antineutrino annihilation 
in the neutrino-transparent, low-density 
plasma before, during, and immediately after the merging of two neutron
stars is not sufficient to explain the energetics of typical gamma-ray
bursts.

\subsection{Black hole formation and accretion}

While the emission of gravitational waves peaks right
at the moment when the two neutron stars fall into each other, the 
neutrino luminosity does not reach a very high level before the tidal
tails have been inflated and wrapped up to the hot gas cloud that 
finally surrounds 
the dense core. Even then, neutrino-antineutrino annihilation is not
an efficient mechanism to provide the energy for a gamma-ray burst,
because electron and positron captures on free nucleons extract the 
deposited energy extremely rapidly from the dense gas that is present
also above the poles of the merger remnant. 

The situation changes, when the core of the remnant collapses to a 
black hole. Our basically Newtonian simulations, however, do not yield
evidence whether and if so, when such a collapse occurs. 
Assuming that it happens, Ruffert \& Janka (1999) 
continued the simulation of Model~B64 for several ms to investigate
the effects on the surrounding gas cloud. The black hole was represented
by a gravitating ``vacuum sphere'', and the time of the gravitational 
instability was treated as a free parameter. The results for the 
dynamical evolution and the neutrino emission, however, turned out to 
be rather similar,
independent of whether the black hole was assumed to form only
$\sim 2\,$ms or as late as 10$\,$ms after the start of Model~B64.

A comparison of Fig.~15 in Ruffert \& Janka (1999) with Fig.~\ref{fig:7}
in the present paper reveals that the baryonic matter above the poles of the 
compact core falls into the newly formed black hole very quickly. Within
milliseconds, a funnel along the rotation axis is cleaned from baryons,
and the density decreases to a value near our numerical lower limit of 
the density. With most of the energy deposition by neutrino-antineutrino
annihilation taking place in that region, this appears to be a 
favorable situation for a baryon-poor, potentially relativistic
outflow to develop, which might later on produce a gamma-ray burst
through dissipation of the mechanical kinetic energy into radiation.

Besides the formation of the black hole, such a scenario requires that
a significant amount of matter remains in a disk around the black hole.
The accretion of this matter on a timescale much longer than the dynamical 
timescale allows for a high efficiency of the conversion of rest-mass 
energy of the accreted gas to neutrinos. Typically several per cent
efficiency in case of a Schwarzschild 
black hole and up to several ten per cent are possible for a Kerr
black hole which accretes matter from a corotating (thin) disk (Thorne~1974,
Shapiro \& Teukolsky 1983, Popham et al.~1999, Li \& Paczy\'nski 2000).
Even more energy is available when the rotational energy of the black hole
is tapped by means of magnetic fields (Blandford \& Znajek~1977, Li~2000,
Lee et al.~2000).

%********************************** fig.14 *****************************************
\begin{figure}
  \epsfxsize=9.2cm \epsfclipon \epsffile{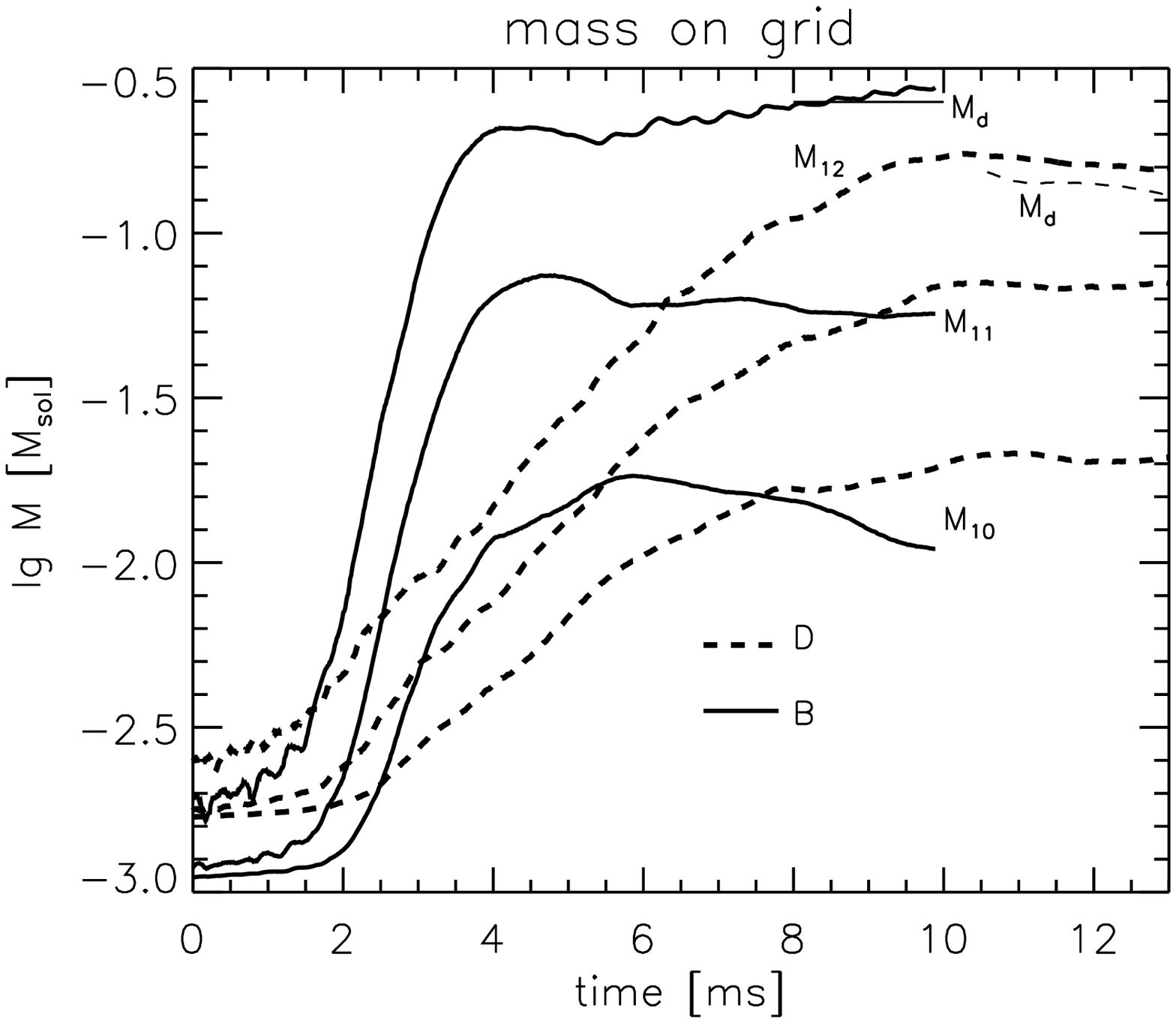}
  \caption[]{Masses on the grid with a density below $10^{10}$~g$\,$cm$^{-3}$
   (``$M_{10}$''), below $10^{11}$~g$\,$cm$^{-3}$ (``$M_{11}$''), and
   below $10^{12}$~g$\,$cm$^{-3}$ (``$M_{12}$''), respectively, for Models~B64
   (solid lines) and D64 (dashed lines) as functions of time. The thin curves
   mark the estimated disk masses, $M_{\mathrm{d}}$,
   which encompass all gas with a specific angular momentum larger than
   the Keplerian angular momentum at three Schwarzschild radii of the merger
   remnant (i.e., of the total mass on the grid)}
\label{fig:14}
\end{figure}
%********************************** fig.14 *****************************************

If a black hole forms from most of
the mass of the merger remnant, our hydrodynamical models allow
us to estimate the mass which ends up in an accretion disk around
this black hole. Assuming that the disk is supported mainly
by centrifugal forces, we use the criterion that the specific angular momentum
of the gas should be larger than the Keplerian angular momentum at three
Schwarzschild radii, i.e., at the location of the innermost stable circular
orbit for a nonrotating black hole: $j > \sqrt{6} GM/c$, where $M$ is taken
to be the total mass on the grid. The corresponding gas masses, $M_{\mathrm{d}}$,
at the end of our simulations are listed in Table~\ref{tab:models1}, and are
indicated for Models~B64 and D64 in Fig.~\ref{fig:14}. Typically, several 
hundredths of a solar mass up to a few tenths of a solar mass fulfill this
condition. The largest numbers are obtained for corotating models (B, S, D),
where the gas has the highest specific angular momentum. The continuation
of the simulation of Model~B64 by Ruffert and Janka (1999) confirmed that
these estimates are in reasonably good agreement with the gas mass which
finally orbits around the black hole when a quasi-stationary state is reached.

For the irrotational A-case (and in the cases where the neutron stars started
out with opposite spin directions or spins in counterrotation to the 
orbit), only a few per cent of the total rest mass of the
binary system can gather enough angular momentum by hydrodynamical
processes to be able to stay in a disk around a black hole 
at the end of our $\sim 10$~ms of computed evolution. These estimates are
confirmed by three-dimensional simulations of neutron star mergers in full 
general relativity (Shibata \& Ury\=u 2000, 2001).
Of course, if the collapse to a black hole occurs immediately after the 
merging of the two neutron stars, before the tidal tails and the extended
envelope of the merger remnant have a chance to form, there is no time for any 
transport of angular momentum by hydrodynamic interaction, and very little
gas, if any at all, will be able to remain in a disk.

On grounds of our current hydrodynamical simulations, solving the 
time-dependent Euler equations, we cannot draw conclusions on the further
development of the accretion ``disk'' or, better, of the extended torus.
Once the gas has settled into a quasi-steady state around the newly-formed
black hole, its later destiny will be driven by the radial transport of 
angular momentum, which is mediated by viscous forces, and the torus structure
and internal conditions will be determined by the energy (and lepton) number loss
through neutrino emission. Magnetic fields can play an important role, too.
In our simulations, without taking into account the effects of the physical disk
viscosity (the numerical value of which cannot be determined from first 
principles and thus would have to be considered as a free
parameter within the Navier-Stokes equations), the subsequent evolution is 
governed by the action of the numerical viscosity, which is not under our
direct control, but can be varied only indirectly by changing the grid 
resolution. In addition, numerical viscosity does not have the same properties
as the physical viscosity, e.g., it does not necessarily conserve the angular
momentum.

Considering the situation at the end of our models we therefore could
only speculate about the long-time evolution of the accretion torus 
(Ruffert \& Janka 1999, Janka et al.~1999). With a given value for the mass
accretion rate by the black hole, for example, we obtained an estimate of the
torus lifetime. Moreover, the mass, density, and temperature of the torus
define its neutrino emission properties, and the calculated neutrino 
luminosity (extrapolated for the estimated accretion time) allows one to
come up with numbers for the efficiencies of conversion of gravitational 
potential energy to neutrinos and to the electron-positron pair plasma by
neutrino-antineutrino annihilation. 

Typical temperatures within the torus are a few MeV, the maximum temperatures
between about 5~MeV and roughly 10~MeV. A significant amount, if not most of
the torus mass has a density above $10^{11}\,$g$\,$cm$^{-3}$ (Fig.~\ref{fig:14}
and Table~\ref{tab:models2}). ``Massive'' tori, i.e., those with a mass of
more than $\sim 0.1\,M_{\odot}$, are optically thick to neutrinos, whereas 
tori with masses of only a few $0.01\,M_{\odot}$ are close to neutrino 
transparency (see Fig.~30 in Ruffert \& Janka 1999). 

Whether the dense core of the merger remnant collapses to a
black hole, and if so, on what timescale this happens, is a complex question,
which requires not only a general relativistic treatment, but depends on
a number of additional aspects, e.g., the mass and compactness of the 
initial neutron stars, the properties of the nuclear EoS,
and the rotation of the post-merging object. Shibata \& Ury\=u (2000, 2001),
by performing three-dimensional dynamical simulations in full general
relativity, find that the product of a neutron star merger is sensitive
to the initial compactness of the neutron stars, which is defined as the ratio
of the Schwarzschild radius of the neutron star to its actual radius. This
quantity increases when the mass of the neutron star approaches the 
limiting mass of a spherical star in isolation. For sufficiently compact 
stars a black hole is formed on a dynamical timescale, as the compactness 
descreases, the formation timescale becomes longer and longer. (The 
corresponding critical 
mass of the binary relative to the mass limit of a nonrotating, single
neutron star varies with the compressibility of the EoS.) For less 
compact cases, a differentially rotating ``supramassive'' neutron star
(Cook et al.~1992, 1994a,b) forms, which first has to become a rigidly
rotating body (on a secular timescale by viscous dissipation) or has 
to lose some of its angular momentum by dynamical processes
(e.g., mass stripping, bar-mode instability) or secular processes
(e.g., gravitational waves, mass loss by winds, magnetic fields), before
the gravitational instability can set in (Shibata et al.~2000).

%********************************** fig.15 *****************************************
\begin{figure}
  \epsfxsize=9.2cm \epsfclipon \epsffile{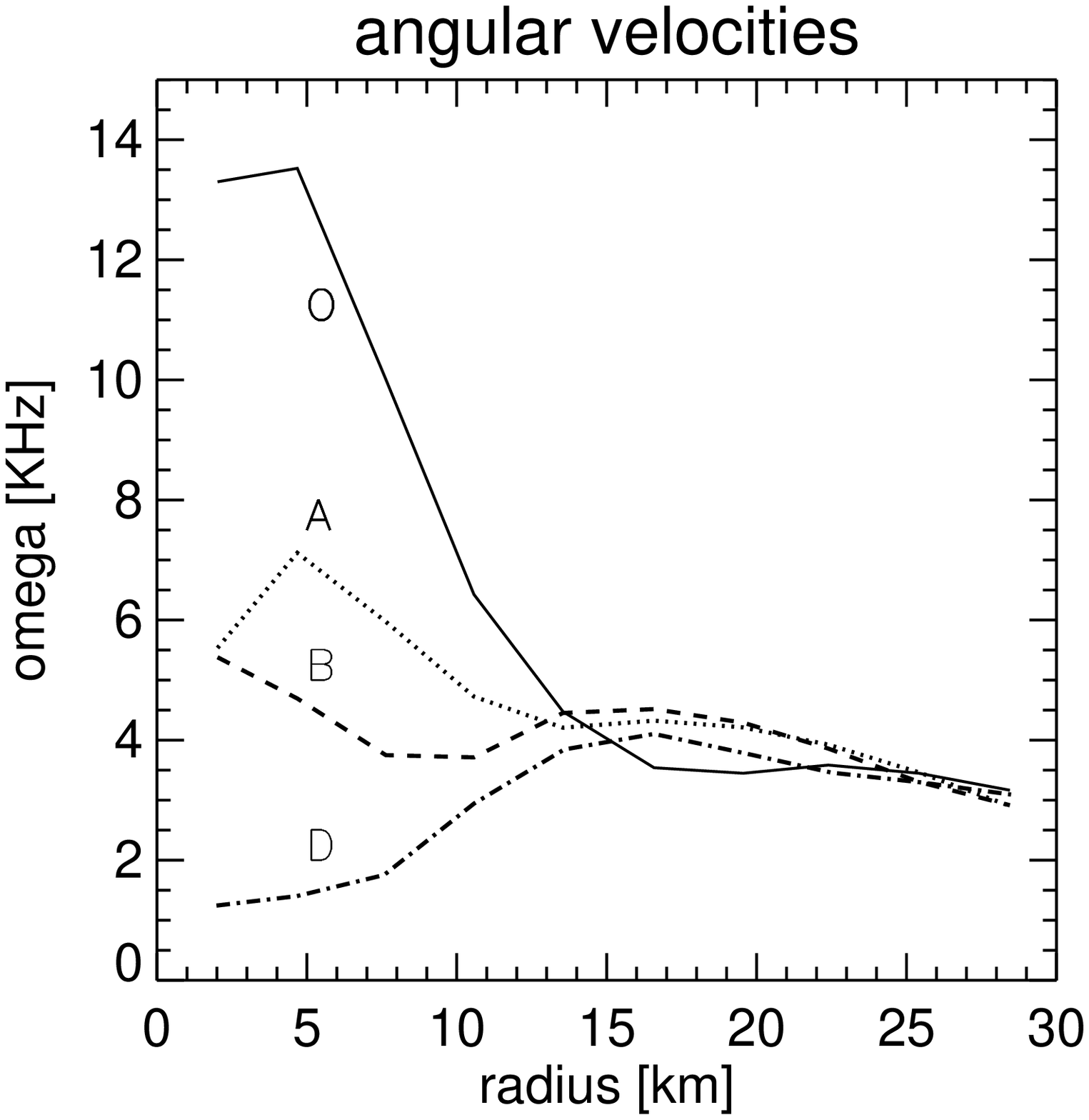}
  \caption[]{Azimuthally averaged angular velocity in the orbital plane
   as a function of the distance from the system axis. The plot shows
   the inner regions of the merger remnants of Models~A64, B64, O64, and
   D64 at the end of the computed evolution
   }
\label{fig:15}
\end{figure}
%********************************** fig.15 *****************************************

Thus the evolution leads to a black hole very quickly only if the total rest
mass of the binary system is sufficiently larger than the maximum rest mass 
of a spherical star in isolation. The exact factor depends on the stiffness
of the EoS as well as on the angular momentum of the binary system. The latter
determines the rotation of the core of the merger remnant, which can remain
stable although it has a mass that is significantly larger than the mass limit 
of spherical or rigidly rotating neutron stars because of the supporting 
effect of rapid, differential rotation (Baumgarte et al.~2000).

Our models have a dimensionless rotation parameter 
$a \equiv Jc/(GM^2)$ ($J$ is the total
angular momentum, $M$ the total rest mass of the system), which initially 
is between 0.64 for Model~C64 and 0.98 for Model~S64. During the evolution, 
the mass changes only slightly, because little gas escapes from the system.
In contrast,
the angular momentum decreases significantly due to the emission of gravitational
waves (and partly due to angular momentum which is carried away by the ejected
gas). The final values of the parameter $a$ are therefore lower, between
about 0.5 for Model~C64 and 0.75 for Model~S64 (see the columns 
$a_{\mathrm{i}}$ and $a_{\mathrm{f}}$ in Table~2 of Janka et al.~1999;
the actual numbers might be up to $\sim 10$\% larger, because one should
apply corrections for the numerical loss of angular momentum).
This suggests that the merger remnant has an angular momentum below the
critical limit that is possible for a Kerr black hole ($a = 1$). 
Of course, a definite statement of this kind is not possible without a 
relativistic treatment, which considers $M$ to be the gravitational mass 
instead of the rest mass, and includes the mass reducing effect of the energy 
loss by gravitational waves. The latter, however, is small. According to the
quadrupole formula less than $\sim 1$\% of the mass of the system is emitted in 
gravitational waves (Fig.~\ref{fig:9a}). Therefore the trend seen in our
Newtonian calculations should be right, and the subcritical rotation of the 
merger remnant should even be quantitatively correct if one does not start
with an initial value of $a$ that is much larger than unity.

It is worth a final remark that the compact core of the merger remnant 
at its ``surface'' near 25~km rotates with an angular velocity that is only 
slightly lower than the Keplerian frequency, 
$\Omega_{\mathrm{K}} = \sqrt{GM/r^3}\sim (4$...$5)\times 10^3\,$s$^{-1}$.
The degree of differential rotation, however, varies strongly between
the different models and depends on the initial neutron star spins and 
the mass ratio of the neutron stars. Figure~\ref{fig:15} shows the azimuthally
averaged angular velocities in the orbital plane for the core of the remnant in
different models at the end of
the computed evolution. Whereas in the irrotational, symmetric case, Model~A64,
the core rotates moderately differentially with radially decreasing mean angular
velocity, the asymmetric Model~D64 reveals the opposite
trend, and the corotating case, Model~B64, is near rigid rotation.
The most dramatic differential rotation is found in the inner
region of Model~O64, where the mean angular velocity drops from roughly
$14\times 10^3\,$s$^{-1}$ near the center to a value around 
$3.5\times 10^3\,$s$^{-1}$ between 15~km and 30~km distance from the system
axis. We would like to emphasize, however, that except for Model~O64, the
different grid zones at a fixed equatorial radius show large fluctuations of 
the corresponding value of $\Omega$, because the merger remnants are still 
significantly perturbed and perform violent oscillations.

\section{Summary, discussion, and conclusions}
\label{sec:conclusions}

We have presented results of neutron star merger simulations that 
were performed with a 
new version of our numerical code, which was significantly improved
compared to the original one (Ruffert et al.~1996, 1997a), mainly 
by the introduction of several levels of nested grids. These allow
for a better resolution of the stars near the grid center on the one
hand, and a larger computational volume on the other. Besides recomputing
our previously published models with the new code, we varied the neutron
star parameters (masses, mass ratios, spins) and computed models with
different maximum resolution. The lower resolution on the finest grid 
was similar to the
grid in our older calculations, whereas the current ``standard models'' 
have a factor of two higher resolution in all three cartesian directions. 
Moreover, we tested the accuracy (or
uncertainty) of the temperature calculation by replacing the energy 
(i.e., the sum of kinetic and internal energies in our code) by
the entropy as the basic variable to follow the temperature
evolution in time.

Using the energy for deriving temperatures has the disadvantage that
the thermal energy is only a small contribution to the internal
energy. The latter is dominated by the degeneracy energy of the fermions. 
This leads to errors in the temperature determination when the 
internal energy is not very accurate. With the entropy equation one also
reduces, although cannot eliminate, the problem that unphysical shocks at
the interface between the neutron stars and the surrounding medium occur
and cause an overestimation of the temperature, in particular before the
merging. Even more, the entropy equation gives one control over the effects 
of shear and bulk viscosity on the thermal evolution, whereas the temperature
calculated from the energy equation does not allow this direct control, 
because the dissipative effects of numerical viscosity in the code can 
be influenced only by changing the grid resolution. Of course,
solving the entropy equation as a supplementary equation in addition to the
conservation laws of mass, momentum, energy, and
lepton number, which still describe the evolution of the stellar fluid
(and where the neutrino source terms were evaluated with the 
temperature as obtained from the entropy equation), 
is not a hydrodynamically fully consistent approach. 
We therefore do not consider this procedure as the necessarily more 
accurate calculation, but as an attempt to test the sensitivity of our
results to effects that are associated with the uncertainties of the
temperature determination discussed above. It is reassuring 
that despite of significant differences of the thermal evolution 
our main results show rather little variation.

In fact, this work was also strongly motivated by the wish to investigate and 
outline (at least some) major uncertainties of (not only our) current neutron
star merger simulations. Such uncertainties have a bearing on the possibility
to draw model-based conclusions on the gravitational-wave emission, a potential
connection with gamma-ray bursts, and the implications for the production of
heavy elements in our Galaxy. The models presented in this paper improve our
previous calculations with respect to numerical resolution and reduced mass
loss through the outer boundaries of the significantly enlarged computational
grid. They are intended to serve for comparisons with future general 
relativistic simulations.

\subsection{Main results}

In summary, our main results and their implications are the following. 
The details of the 
gravitational-wave signal, i.e., the primary and subsequent maxima of the 
luminosity, the total radiated energy, and the 
structure of the wave amplitude in particular 
during the post-merging phase, when the core of the merger remnant is in a 
highly perturbed state and performs violent oscillations, exhibit some change
with the resolution on the finest grid. A cell size of 0.64~km or smaller, 
corresponding to at least 50 zones on the diameter of the initial neutron
star, seems desirable.
With less resolution, the effects of numerical viscosity and the
associated loss of angular momentum grow to an unacceptable level
and characteristic features of the gravitational-wave emission are
noticably influenced. This limits the possibility to deduce important
information from a possible future wave measurement. The
gravitational waves which accompany the final plunge of the 
neutron stars, carry information about the masses, compactness and the
spins of the neutron stars and thus allow for conclusions on the nuclear EoS.
The post-merging signal yields evidence about the 
destiny of the merger remnant and can also be used to extract information
about the internal state of neutron stars.

While the gravitational-wave emission is strongest when the two neutron stars
plunge into each other, the neutrino emission rises only gradually afterwards.
It approaches a saturation level towards the end of our simulations, when the
tidal tails have been wrapped up to a cloud of shock- (and shear-)heated gas
that surrounds the compact and much denser core of the merger remnant.
Because of the more extended computational grid, which is necessary to 
follow the development of this gas cloud, the neutrino luminosities are found
to be a factor of 2--4 higher than in our previously published models  
(Ruffert et al.~1997a). Correspondingly, the energy deposition rates 
by neutrino-antineutrino annihilation in the
dilute outer layers ($\rho \la 10^{11}\,$g$\,$cm$^{-3}$) of the 
post-merging object are up to a factor of 20--30 larger. By far the dominant
part of this energy, however, is deposited in the polar regions, where the
temperature is high and the scattering depth to neutrinos is rather low. 
Therefore this 
energy is immediately reradiated through neutrinos produced by electron and 
positron captures on nucleons. Such an energy transfer to a region with large 
baryon density is therefore inefficient in powering a gamma-ray burst, and 
should drive a baryonic wind rather than a relativistic outflow of 
a baryon-poor pair-photon plasma. This presumably high-entropy wind
(i.e., the medium has low density but comparatively high temperature),
which expands into vacuum,
may have very interesting, so far not investigated, implications
for observable radiation from neutron star merging events, and for
the enrichment of our Galaxy with heavy elements 
(a more detailed discussion can be found in Ruffert \& Janka~1999 and 
Janka \& Ruffert~2001). 

The dynamical mass ejection from the merging binary varies with the initial
spins of the neutron stars. It is largest ($\sim (2...4)\times 10^{-2}\,M_{\odot}$
or roughly 1\% of the system mass) for corotating systems (a case
which is not likely to be realized because of the small viscosity of    
neutron star matter, which cannot bring the system into a tidally locked
state prior to merging; Kochanek 1992, Bildsten \& Cutler~1992) and smallest 
($\la 10^{-4}\,M_{\odot}$) when the stars initially counterrotate with the orbit.
Only in the former case very prominent tidal arms develop through the outer 
Lagrange points and expand away from
the center of the merger. The use of the larger
computational volume helped to significantly improve our estimates for the 
amount of mass which can potentially become unbound. 

It is an important aim of numerical models to determine the thermodynamical
conditions and the nuclear composition of these ejecta, and their subsequent
evolution. This will help answer the question whether and how r-processing 
can take place in this matter (Lattimer \& Schramm 1974, 1976;
Symbalisty \& Schramm 1982; Meyer 1989;
Davies et al.~1994; Freiburghaus et al.~1999) and whether it has contributed
to the heavy-element content of our Galaxy at a significant level. 

We find that the matter, which is ejected from the tips of the
expanding tidal tails, stays cool, because it is neither heated
by shocks nor by viscous friction. In fact, it is a problem in hydrodynamical
simulations to accurately trace the thermal history of the initially cold 
neutron matter (the viscosity is not only too small to achieve tidal locking, it 
is also too small to heat the neutron stars beyond $\sim 10^9$~K before merging;
Lai 1994). Besides the high degeneracy of the medium,
numerical viscosity, which is present to some
(but actually different) degree in all numerical codes and depends on the
resolution, causes problems for an accurate calculation of the temperature. 
In addition, the limited numerical resolution does not allow one to 
properly represent the extremely steep density gradient near the neutron
star surface below a density of about $10^{14}\,$g$\,$cm$^{-3}$.
This leads to the necessity of softening the density gradient
to produce (nearly) hydrostatic conditions.
In our Eulerian, grid-based simulations the neutron stars also have to be
embedded by a low-density medium. When the neutron stars move through
this surrounding medium, shocks occur at the stellar surfaces in upstream
direction, which produces artificial heating. Thus locally, the temperatures
can be overestimated by a large factor, although this numerical heating is
small compared to the maximum temperatures which are reached when the 
two neutron stars plunge into each other. Because of all these aspects,
some of which are not specific to a particular code but are generic 
to the physical problem or to hydrodynamical simulations with limited 
resolution, the calculated temperatures are likely to be overestimated 
and have to be interpreted with special caution.
Significant progress requires a much 
enlarged numerical resolution. For grid-based codes this could be
achieved by employing adaptive mesh refinement techniques. Another
option for improving some aspects may be the choice of a rotating
instead of a fixed frame of reference (New \& Tohline 1997,
Swesty et al.~2000), although the inspiral is so rapid that there
would quickly be motion of the stars also in a reference frame that 
is initially corotating.

Our merger simulations followed the variation of $Y_e$ in response to
electron captures by protons and positron captures by neutrons, 
including also the effects of neutrino trapping when the stellar medium 
becomes optically thick to neutrinos at sufficiently large densities.
Although these weak interaction rates are fast at temperatures
between $10^{10}\,$K and $10^{11}\,$K, they are not effective in raising 
the initial electron fraction on the very short timescale of the dynamical
expansion of the ejecta. Starting with a typical equilibrium $Y_e$ around 
0.02 in the tidal tails, we cannot find a growth by more than a factor 
of 2--3 (to values of at most $\sim 0.06$) before the gas leaves our 
computational grid. At that time the density has decreased to 
$10^{10}\,$g$\,$cm$^{-3}$ or less, and the matter has cooled down to
(2--$3)\times 10^{10}\,$K. Since both the temperature and the density drop
quickly, and more and more nucleons recombine to alpha particles and nuclei,
we do not expect a significant effect due to electron and 
positron captures during the subsequent expansion. This result was obtained
although the temperature (and therefore the mass fraction of free nucleons)
was overestimated in our simulations, implying unrealistically fast electron
and positron captures on the free protons and neutrons. In comparative
calculations, using the entropy equation for following the thermal history of
the medium (see above) and starting with lower (and thus more realistic)
temperatures, we actually cannot detect any change from the initial value 
of $Y_e$.
We emphasize that our point here is the fact that neutrino processes
seem to be unable to alter $Y_e$ during the rapid decompression of essentially
cold neutron star matter. The initial $Y_e$ of the neutron star crust is
therefore preserved, although the exact value may be EoS dependent and is
therefore uncertain.

\subsection{Elements of a possible r-process site}

Cold material in neutron star crusts consists of neutron-saturated, very 
neutron-rich nuclei that are arranged on a lattice and immersed in a
gas of neutrons and degenerate electrons (e.g., Weber 1999). 
This region has a very
low proton-to-nucleon ratio ($Y_e \sim 0.02$; only in a thin skin layer of
the neutron star, the outer crust, the electron fraction rises again).
As discussed above, it is unlikely that the decompressed and expanding 
matter in the tidal tails is heated by dissipative processes (shocks,
viscous shear) to temperatures where nuclear 
statistical equilibrium sets in ($T \ga 5\times 10^9$~K). Therefore the
memory of the initial nuclear composition is not erased during this
phase of the expansion, and the subsequent changes of
the nuclear abundances have to be determined by following the
beta-decays of the initial ensemble of heavy nuclei (Hilf et al.~1974,
Lattimer et al.~1977, Meyer 1989). Since heating by beta-decays can
raise the temperatures to several $10^9$~K without, however, necessarily
destroying the memory of the initial composition (Meyer 1989, 
Sumiyoshi et al.~1998), a self-consistent coupling of the hydrodynamical
evolution with the effects of nuclear decays, 
including the decay heating and possible $(\gamma,p)$ and 
$(\gamma,n)$ reactions, is essential for
reliable predictions of the final nucleosynthetic yields. 
Calculations fulfilling these conditions have not been performed so far,
and it remains to be seen whether this scenario yields a solar system
like distribution of r-process elements.

In contrast, Freiburghaus et al. (1999) assumed that the ejected gas
starts from nuclear statistical equilibrium, i.e. with a temperature 
above $\sim 5\times 10^9\,$K, because they used the EoS of
Lattimer \& Swesty (1991), which was actually
developed for supernova conditions and does not contain the physics
required to describe cold neutron stars. Combining hydrodynamic results of
neutron star merger simulations with network calculations, they found 
that for proton-to-nucleon ratios around 0.1 rapid neutron captures 
produce an abundance pattern which fits the observed solar r-process
very well for nuclear masses around and above the $A\approx 130$ peak.
However, they considered $Y_e$ as a free parameter instead of determining
it as a result of neutrino emission and absorption reactions in the 
hydrodynamical merger model. Moreover, the feedback of beta-decay heating 
on the hydrodynamic evolution of the fluid elements was not taken into 
account.

Based on our simulations we come to the
conclusion that the ejected gas stays cool, does not get heated by shocks or
viscous shear to temperatures where nuclear statistical equilibrium holds,
and retains its very low initial proton-to-nucleon ratio. Therefore our
simulations do not yield the conditions which were determined
by Freiburghaus et al.~(1999) as favorable for producing a solar-like 
r-process abundance pattern in the $A\ga 130$ mass range. Our models do
not provide evidence that their values for the initial temperature and 
composition are realised in the ejecta from neutron star mergers. Instead,
our results seem to favor the kind of scenario discussed by Lattimer et al.~(1977)
and Meyer~(1989), who considered the decompression of initially cold
matter from the inner crust of neutron stars, where the composition is
dominated by extremely neutron-rich nuclei (heavy metals) that can be
arranged on a lattice and are immersed in a gas of neutrons and relativistic
electrons. However, it is unclear whether the decompression from such
an initial state leads to the abundance distribution of rapid neutron 
capture elements as observed in the solar system.

Besides beta-decays, electron and positron captures and $(\gamma,p)$ and
$(\gamma,n)$ reactions,
a detailed discussion of the nucleosynthesis in the dynamically ejected 
matter might also require taking into account the
interaction with the intense fluxes of energetic neutrinos from the 
merger remnant. Neutrinos absorbed by nucleons and scattered by nuclei in the
outflowing gas may heat and accelerate the gas, may change the 
proton-to-neutron
ratio and may reprocess the heavy elements by inducing nuclear
transmutations. Although the neutrino emission from the remnant rises 
only gradually after the tidal tails have formed,
and the timescale of the expansion of the gas away from
the merger site is very short (of the order of the escape timescale, which
is roughly $t_{\mathrm{exp}}\sim r/v_{\mathrm{esc}}\sim 1\,$ms), the 
number of neutrino-nucleon interactions can be estimated to be
significant. An order of magnitude evaluation shows that about 
10\% of the nucleons might react with neutrinos:
\begin{equation}
\int_{R_i}^\infty {\mathrm{d}}r\,{{\cal R}\over
n_{\mathrm{b}}v(r)}\,\sim\,0.1\,\alpha\,
     \ave{\epsilon_{\nu}}_{20}\, L_{\nu,53}\, (M_3\, R_{i,7})^{-1/2} \, ,
\end{equation}
where ${\cal R}$ is the reaction rate per neutrino per unit of volume,
$\ave{\epsilon_{\nu}}_{20}$ the mean energy of the emitted neutrinos 
measured in 
20~MeV, $L_{\nu,53}$ the total neutrino luminosity in $10^{53}\,$erg$\,$s$^{-1}$,
$M_3$ the mass of the remnant normalized to $3\,M_{\odot}$, and 
$R_{i,7}$ the initial radius in units of $10^7\,$cm. The radial velocity
was assumed to be roughly given by $v(r) = \sqrt{2GM/r}$, and the 
neutrino interaction cross section was approximated by the absorption
cross section of $\nu_e$ and $\bar\nu_e$ on nucleons,
$\sigma \sim 10^{-43}(\epsilon_{\nu}/1\,{\mathrm{MeV}})^2\,$cm$^2$, because
electron neutrinos and antineutrinos dominate the neutrino emission of the
merger remnant. The quantity $\alpha$ is a geometrical factor and is 
less than unity. It accounts for
the fact that the neutrinos are radiated mainly perpendicular to the 
orbital plane (because in this direction the scattering depth
is smaller), whereas the dynamically ejected gas moves away 
from the system in the plane of the binary orbit.

The amount of dynamically ejected material varies strongly with the 
neutron star spins and is largest for the improbable case of corotating
systems. Moreover, it is sensitive to the stiffness of the unknown nuclear EoS
(Freiburghaus et al.~1999, Rosswog et al.~2000). Because momentum has
to be transferred by hydrodynamical processes
before gas can be expelled, mass ejection may even
be suppressed by a quick collapse of the merger remnant to a black hole,
a possibility which depends on the neutron star
equation of state, the masses of the merging stars, and the angular momentum 
of the binary system (Shibata \& Ury\=u~2001). In view of these fundamental
uncertainties, current models are unable to yield quantitatively
meaningful numbers for the contribution of neutron star mergers to the 
metal enrichment of our Galaxy, even if the theoretical
estimates of the merger rates were reliable (which in fact is not the case).

\subsection{Concluding remarks}

We have presented results from state-of-the-art simulations of 
neutron star mergings with the most advanced version of our hydrodynamics
code, and have discussed these results and their limitations.
Of course, our calculations are far from being satisfactory.
Relativistic simulations are necessary to address the question
whether a black hole forms and if so, on what timescale it happens. 
This is important not only for predictions of the gravitational-wave signal 
and the mass which can become unbound during the merging. It is also 
important for studying the implications of neutron star mergers for the
origin of heavy elements and has consequences for potentially 
observable signals connected with the neutrino- and photon-cooling phases
of the remnant, which is either a rotating (supramassive), hot neutron 
star or a black
hole which accretes matter from a surrounding torus at very high rates.

So far our treatment of neutrino production and emission by using a trapping 
scheme does not allow us to study the effects of neutrino transport and 
neutrino energy deposition in the outer layers of the merger remnant. The
latter should drive a baryonic outflow from the massive neutron 
star or from the accretion disk around the black hole. Including the 
feedback of neutrino interactions is essential for modeling this mass loss
and for determining the conditions in the wind. 

This discussion shows that the modeling of the merging of neutron stars
and of the accompanying physical processes is still at its beginning. 
Current simulations do neither fully account for the effects of 
general relativity, nor do they include all the physics which is relevant
to come up with meaningful predictions of the potentially observable photon 
and neutrino emission, or to clarify the role of neutron star mergers for the 
production of heavy elements in our Galaxy. Therefore conclusions
drawn from current hydrodynamical models have to be considered with 
special reservation. We have outlined a number of aspects where improvements
and progress in future simulations are highly desirable.

\begin{acknowledgements}
It is a pleasure for us to thank Wolfgang Keil for transforming
Lattimer \& Swesty's FORTRAN equation of state into a usable table.
We would also like to thank an anonymous referee for his careful
reading and useful comments. MR is grateful for support by a PPARC
Advanced Fellowship, HTJ acknowledges support by the
``Sonderforschungsbereich 375 f\"ur Astro-Teilchenphysik''
der Deut\-schen For\-schungs\-ge\-mein\-schaft.
The calculations were performed at the Rechenzentrum Garching on a
Cray-YMP 4/64 and a Cray-EL98~4/256.
\end{acknowledgements}

\end{document}